\documentclass[10pt,a4paper]{article}
\usepackage{jheppub_kim}
\usepackage{cleveref}
\usepackage{pdflscape}
\usepackage{amsmath}
\usepackage{amssymb}
\usepackage{dcolumn}
\usepackage{bm}
\usepackage{color}
\usepackage{amsmath}
\usepackage{breqn}
\usepackage{graphicx}
\usepackage{epsfig}
\usepackage{caption,subcaption}
\usepackage{amsfonts}
\usepackage{graphicx}
\usepackage{parskip}
\usepackage{dcolumn}
\UseRawInputEncoding

\begin{document}

\title{New Agegraphic Dark Energy in Braneworld Cosmology: A Higher Dimensional Perspective on Dark Energy Evolution}

\author[a]{Pameli Saha}
\author[b]{Aritra Sanyal}
\author[b]{Prabir Rudra}

\affiliation[a] {Department of Mathematics, Amity School of Applied Sciences, Amity University, Maharashtra-410206, India}

\affiliation[b] {Department of Mathematics, Jadavpur University, Kolkata-700 032, India.}

\affiliation[b] {Department of Mathematics, Asutosh College,
Kolkata-700 026, India}

\emailAdd{pameli.saha15@gmail.com}
\emailAdd{aritrasanyal1@gmail.com}
\emailAdd{prudra.math@gmail.com}

\abstract{In this study, we investigate the new agegraphic dark energy model within the context of braneworld cosmology. A higher dimensional viewpoint on the evolution of dark energy is examined. A combination of cold dark matter and dark energy, is examined within the frameworks of RS-II and DGP brane as the underlying gravitational theories. Both the possibilities involving interaction and non-interaction between dark energy and cold dark matter are investigated. A range of cosmological parameters, including the equation of state parameter, deceleration parameter, and statefinder parameters, are examined. The squared velocity of sound is examined to assess the system's stability. A data analysis based on observations, utilizing contemporary cosmological data and the Markov chain Monte Carlo method, is conducted to constrain the parameter space of the model. Our research suggests that the interplay between higher-dimensional phenomena and agegraphic dark energy provides a theoretically robust and observationally consistent framework for the universe's late-time acceleration.}

\maketitle

\section{Introduction}
\label{sec:intro}
Substantial and reliable observational data from various sources have led to a significant change in cosmology in recent years, demonstrating that the universe is expanding at an accelerating rate \cite{acc1, acc2, acc3}. Diverse and independent datasets, such as observations of Type Ia supernovae \cite{acc1, acc2, acc3}, the large-scale matter distribution, temperature fluctuations in the Cosmic Microwave Background Radiation (CMBR), data from the WMAP mission \cite{wmap1, wmap2}, and measurements of Baryon Acoustic Oscillations (BAO) \cite{bao1, bao2, bao3}.
According to these observations, dark matter (DM) and dark energy (DE) are the two mysterious components that dominate the cosmos \cite{de1, de2}. DM contributes to the explanation of galaxy rotation curves and the creation of cosmic structures, while DE is responsible for the universe's current accelerating expansion. The energy-momentum tensor (EMT), which is found on the right side of Einstein's field equations, is frequently altered in DE models to account for the rapid expansion. Modified gravity theories, on the other hand, concentrate on altering the geometry, or the left side of these equations. Notwithstanding these strategies, a comprehensive theory that can concurrently account for the universe's accelerated expansion, early-universe problems, structure formation, DM, and other basic challenges is still unattainable. Incorporating quantum effects into gravity is a critical step toward such a theory. Although a full theory has not yet been developed, quantum gravity seeks to combine the ideas of quantum mechanics and general relativity. The Holographic dark energy (HDE) \cite{hde1, hde2, Rudra2023a} and Agegraphic dark energy (ADE) \cite{nade1, nade2} models stand out among the many quantum gravity-based theories as possible causes of the universe's late-time acceleration. The cosmological constant problem \cite{cc1, cc2, cc3, cc4, cc5} and the cosmic coincidence problem \cite{cc3, cc4, cc5} are two significant cosmological challenges. Several modified gravity theories have been used to study these issues \cite{pr1, pr2, pr3}.

The ADE framework originates from quantum mechanics, namely the uncertainty principle, and incorporates gravitational influences from general relativity (GR). It elucidates dark energy via alterations in spacetime and the composition of matter as dictated by the universe. Drawing on the concept of quantum fluctuations inside space-time, Karolyhazy and his associates \cite{karo1, karo2, karo3} suggested that in Minkowski space-time, the assessment of a time interval $t$ cannot achieve a precision superior than 
\begin{equation}
\delta t=\gamma t_p^{\frac{2}{3}} t^{\frac{1}{3}},
\end{equation}
where $t_p$ is the reduced Planck time and $\gamma$ is a dimensionless constant of order one. Maziashvili \cite{karo4} proposed that the quantum energy density resulting from metric fluctuations in Minkowski space-time can be approximated by utilizing this uncertainty relation alongside the time-energy uncertainty principle. Both Maziashvili \cite{karo4} and Sasakura \cite{karo5} independently formulated the equation for the energy density of these variations as
\begin{equation}
\rho_D \sim \frac{1}{t_p^2 t^2} \sim \frac{M_P^2}{t^2}
\end{equation}
where $M_P^2=(8\pi G)^{-1}$ is the reduced Planck mass. This energy density is understood as dark energy, particularly within the framework of the Agegraphic Dark Energy (ADE) paradigm. The initial ADE model, proposed by Cai \cite{nade1}, attempted to understand the universe's accelerating expansion. It defines the energy density as
\begin{equation}
\rho_D = 3n^2M_P^2T^{-2}
\end{equation}
where $T$ represents the age of the universe, $M_P$ is the Planck mass, and $n$ accounts for certain uncertainties. Notwithstanding its sophistication, the original ADE model has considerable dynamical issues, especially during the primordial cosmic epoch. The universe possesses a brief lifespan $T$ during the matter-dominated epoch. Consequently, $\rho_{D}\sim 1/T^{2}$ increases significantly. Consequently, dark energy would dominate prematurely, contradicting the necessary matter-dominated epoch essential for structure formation. In summary, ADE cannot authentically replicate the correct sequence of cosmic evolution. Moreover, the ADE model forecasts an erroneous scaling of dark energy in relation to the scale factor $a$.  Discrepancies between observational data and structural growth arise from the equation of state $\omega_{D}$ remaining excessively close to zero in the past. The model exhibits non-locality in time when $T$ is characterized as the total age of the universe, as the energy density relies on the entirety of the universe's historical timeline, which is somewhat unphysical in the context of dynamic cosmology.

To tackle this issue, Wei and Cai \cite{nd1} proposed the new Agegraphic dark energy (NADE) model, which replaces the age of the universe with conformal time $\eta$. This updated model inherently addresses the problem of coincidence \cite{nd2}. A significant benefit of the NADE model is that it possesses an equivalent number of parameters as the $\Lambda$CDM model (which incorporates the cosmological constant and cold dark matter), rendering it more efficient than alternative dynamical dark energy models. Furthermore, the NADE model has shown strong concordance with empirical data \cite{nd3, nd4, nd5}. In \cite{nda1}, models of entropy-corrected holographic and new agegraphic dark energy are examined within the context of generalized Rastall gravity, utilizing two distinct scale factors: one indicative of a future singularity and the other of an initial singularity. The occurrence of accelerated expansion in a flat D-dimensional fractal Universe has been investigated through the analysis of non-linear interactions involving cold dark matter and several DE models, such as Tsallis, Renyi, and Sharma-Mittal HDE and NADE as referenced in \cite{nda2}. In \cite{nda3}, the authors examined the thermodynamic principles regulating the interplay between NADE and DM within an anisotropic universe, deriving the relevant expressions for the entropy changes of these DE constituents. In \cite{nda4}, a novel ADE model is developed inside the $f(Q)$ gravity paradigm, presuming a power-law scale factor in a flat Friedmann-Robertson-Walker (FRW) spacetime. The examination of the $r-s$ plane, coupled with stability assessments, indicates a quintessence phase and demonstrates that the model proficiently resolves the cosmic coincidence issue. In \cite{nda5}, a novel ADE model is examined within the context of $f(Q,T)$ gravity in a flat FRW universe characterized by a power-law scale factor. The model demonstrates a quintessential phase that aligns with the characteristics of Chaplygin gas in the $r-s$ plane. In \cite{nda6}, an extensive global fitting analysis of the NADE model in a non-flat universe is conducted utilizing the Markov chain Monte Carlo (MCMC) technique, which integrates the complete CMB power spectrum from WMAP 7-year data, SNIa observations from the Union 2.1 dataset, BAO measurements from SDSS DR7 and WiggleZ surveys, along with the most recent $H_{0}$ estimates from HST. Other notable work in NADE can be found in refs. \cite{nd6, nd7, nd8, nd9, nd10}. 

The higher-dimensional framework offered by braneworld cosmology depicts the observable Universe as a four-dimensional brane immersed in a higher-dimensional bulk spacetime \cite{rs1, rs2, rs3}. Ordinary matter and fields are often restricted to the brane in such models, although gravity can travel through the extra dimensions. Generalized Friedmann equations with extra energy-density and geometrical corrections result from the interaction between the brane and the bulk, which alters the effective gravitational dynamics on the brane \cite{rs4, rs5, rs6, rs7}. Novel cosmic phenomena, including modified early-Universe dynamics, late-time accelerated expansion, and potential explanations for dark energy and other cosmological pressures, can result from these adjustments \cite{dgp1, dgp2}. As a result, braneworld scenarios offer a helpful framework for examining deviations from conventional four-dimensional cosmology and probing potential extra-dimensional fingerprints.

The observed late-time accelerated expansion of the Universe continues to extend a significant issue in contemporary cosmology, prompting the investigation of dark energy theories that extend beyond the conventional cosmological constant framework. Among the diverse dynamical dark-energy models, NADE stands out because to its energy density being derived from the Universe's conformal age and its strong association with quantum fluctuations of spacetime \cite{nd1, nd2}. Simultaneously, braneworld cosmology offers an inherent higher-dimensional context wherein the effective gravitational behavior in our four-dimensional Universe may diverge from the forecasts of conventional GR \cite{rs3}. The modified Friedmann dynamics and high-energy adjustments linked to braneworld models can profoundly affect the evolution of DE, particularly during the early and middle stages of cosmic development. Integrating NADE with braneworld gravity is a chance to explore how the geometric implications of an additional dimension alter the evolution and dynamic characteristics of ADE. Driven by these factors, the current study investigates NADE within a braneworld context, focusing on the ensuing cosmological development, dynamic characteristics, and the influence of higher-dimensional gravitational modifications on the late-time accelerated expansion of the Universe. The basic motivation for exploring NADE within a braneworld context is from the potential impact of higher-dimensional gravitational dynamics on the emergence of a quantum-geometrically inspired DE component. The paper is organized as follows: In section \ref{sec:brane}, cosmological braneworld models are discussed. In section \ref{sec:nade_brane}, the NADE model is set up in the RS-II and DGP braneworld backgrounds for both the non-interacting and interacting cases. In section \ref{sec:params}, the equation of state parameter, the deceleration parameter, the squared speed of sound and the statefinder parameters are analysed. In section \ref{sec:data}, the model parameters are constrained using observational data. In section \ref{sec:hubble_tension}, the Hubble tension is analysed, and in section \ref{sec:conclusion} we present our conclusions.

\section{Cosmological Braneworld models}
\label{sec:brane}
Here we will restrict our cosmological study to two well-known braneworld models, namely, RS II brane \cite{rs1, rs2} and DGP brane \cite{dgp1, dgp2}. Below we discuss them separately in detail. 

\subsection{RS II brane}
\label{subsec:rs2}
The RS II brane model \cite{rs1, rs2} is a bulk-brane model that Randall and Sundrum suggested to describe a higher dimensional theory. This paradigm states that we inhabit a four-dimensional world (a 3-brane, and a domain wall) immersed in a five-dimensional anti-de-Sitter ($AdS_5$) bulk space time. While gravity may only go through the bulk, all matter fields are contained within the brane. This brane model has recently gained acceptance in the field of cosmology due to its consistency with the expanding universe \cite{rs3}. In contrast to traditional four-dimensional cosmology, the geometry of the extra dimension and the finite brane tension $\lambda$ have an impact on the gravitational dynamics in this scenario. The bulk Weyl tensor and the quadratic energy-momentum contribution of the brane matter introduce corrections to the effective four-dimensional Einstein equations on the brane \cite{rs5}.

We consider a flat homogeneous and isotropic FLRW spacetime, 
\begin{equation}\label{flrwmet}
ds^{2}=-c^{2}dt^{2}+a^{2}(t)\left(dx^{2}+dy^{2}+dz^{2}\right)    
\end{equation}
where $a(t)$ is the cosmological scale factor of the expanding universe. Using this, the modified FLRW equation for RS II brane is given by \cite{rs1, rs2, rs3, rs4, rs5, rs6, rs7}
\begin{equation}
H^{2}=\frac{\kappa_{4}^{2}}{3}\rho\left(1+\frac{\rho}{2\lambda}\right)+\frac{\Lambda_{4}}{3}+\frac{\mathcal{U}}{a^{4}}
\end{equation}
where $\kappa_4$ is the four-dimensional gravitational coupling constant, $\rho$ is the matter energy density, $\Lambda_4$ is the effective four-dimensional cosmological constant and $\mathcal{U}$ represents contribution from dark radiation. Moreover, $H=\frac{\dot{a}}{a}$ is the Hubble parameter. Here will be neglect the contributions from the effective four-dimensional cosmological constant and the dark radiation. So our working equation will be
\begin{equation}\label{rs2frw}
H^{2}=\frac{\kappa_{4}^{2}}{3}\rho\left(1+\frac{\rho}{2\lambda}\right)
\end{equation}
The term $\rho^{2}/2\lambda$ becomes significant at high energies, leading to departures from the standard Friedmann dynamics, whereas the conventional four-dimensional cosmology is recovered in the low-energy limit $\rho<<\lambda$ \cite{rs3}. As a result, the RS-II framework offers an ideal environment for investigating high-energy cosmological impacts and early-Universe dynamics alterations brought about by the existence of an additional spatial dimension.

\subsection{DGP brane}
\label{subsec:dgp}
The Dvali-Gabadadze-Porrati (DGP) braneworld model \cite{dgp1}, which represents our four-dimensional world as a FLRW brane nested in a five-dimensional Minkowski bulk, is a straightforward and useful model of brane gravity. It describes the origin of dark energy as the large scale leakage of gravity from the brane to the bulk \cite{dgp2}. The action of gravity is proportional to the Planck mass $M_{p}^{2}$ on a 4-dimensional brane and to the corresponding quantity in 5 dimensions in the bulk. The model is then characterized by a cross over length scale $r_{c}=\frac{M_{4}^{2}}{2M_{5}^{2}}$ (where $M_{4}$ and $M_{5}$ denote the four and five-dimensional Planck masses, respectively) such that gravity is 4-dimensional theory at scales $a<<r_{c}$, where matter behaves as pressure-less dust, but gravity leaks out into the bulk at scales $a>>r_{c}$ and matter approaches the behavior of a cosmological constant \cite{dgp1, dgp2}. Furthermore, it has been demonstrated that the DGP brane can firmly embed the ordinary Friedmann cosmology \cite{dgp2}. The typical DGP model has been generalized in the literature to (i) LDGP model by adding a cosmological constant \cite{dg1}, (ii) QDGP model by adding a quintessence perfect fluid \cite{dg2}, (iii) CDGP model by Chaplygin gas \cite{dg3}, and (iv) SDGP by a scalar field \cite{dg4}.

Considering a flat homogeneous and isotropic FLRW spacetime (\ref{flrwmet}), the modified FLRW equations for DGP brane \cite{dgp2, rs8} is given by
\begin{equation}\label{dgpfrw}
H^{2}+\frac{\epsilon}{r_{c}}H=\frac{\kappa_{4}^{2}}{3}\rho
\end{equation}
where $\epsilon=\pm 1$ (choosing $M_{p}^{2}=8\pi G=1$) is a constant.

In cosmology, the DGP scenario modifies the usual Friedmann equation and can generate late-time cosmic acceleration without the need for an explicit cosmological constant in its self-accelerating branch \cite{dgp2}. In contrast, the normal branch typically needs an extra dark-energy component to produce acceleration \cite{dg1, dg2}. As a result, the DGP brane model offers an intriguing geometrical modification of gravity as well as a helpful framework for examining potential departures from ordinary four-dimensional general relativity and late-time cosmic acceleration.

\section{NADE in Braneworld Cosmology}
\label{sec:nade_brane}
Here, we analyze the basic setup that entails an interaction between cold dark matter (CDM) and the NADE within a braneworld setting. The energy density associated with the NADE model is specified by \cite{nd1}
\begin{equation}\label{nade}
\rho_{D}=3n^{2}M_{p}^{2}\eta^{-2}
\end{equation}
where $n$ is an uncertainty parameter and $\eta$ is the conformal time given by
\begin{equation}\label{NON2}
\eta=\int_0^t \frac{dt}{a}=\int_0^a \frac{da}{Ha^2}
\end{equation}
Now we will consider two different case depending on the nature of interaction between NADE and CDM. There can be a case where there is no interaction between the two sectors, and there can be a case where the two components interact with each other. We study the two cases separately.

\subsection{Non-interacting Case}
If we assume that NADE and CDM do not interact and are conserved independently, the corresponding conservation equations for the components yield
\begin{equation}\label{dmcon}
\dot{\rho}_{m}+3H\rho_{m}=0,
\end{equation}
and
\begin{equation}
\dot{\rho}_{D}+3H\rho_{D}(1+\omega_{D})=0.
\end{equation}
where $\rho_{m}$ is the energy density of CDM, and $\omega_{D}=p_{D}/\rho_{D}$ is the equation of state (EoS) parameter of NADE. Solving eqn.(\ref{dmcon}) we get the density of matter as
\begin{equation}\label{denmat}
\rho_{m}=\rho_{m_{0}}(1+z)^{3}
\end{equation}
where $\rho_{m0}$ is the current energy density of CDM.

Using eqns.(\ref{rs2frw}), (\ref{nade}), and (\ref{denmat}) the FRW equations of RS-II brane with NADE and CDM can be obtained as,
\begin{equation}
H(z)=\frac{1}{\sqrt{3}}\sqrt{\frac{1}{\lambda^{2}M_{p}^{4}\Omega_{D}^{2}}\left[\lambda (M_{p}^{2}\Omega_{D}-1)+3H_{0}^{2}M_{p}^{4}\Omega_{D}\Omega_{m_{0}}(1+3z+3z^{2}+z^{3})+P\right]}
\end{equation}
where
\begin{equation*}
P=\sqrt{6H_{0}^{2}\lambda^{3}M_{p}^{6}\Omega_{D}^{2}\Omega_{m_{0}}(1+z)^{3}+6H_{0}^{2}\lambda M_{p}^{4}\Omega_{D}(-1+M_{p}^{2}\Omega_{D})\Omega_{m_{0}}(1+z)^{3}-Q}
\end{equation*}
and
\begin{equation*}
Q=-9H_{0}^{4}M_{p}^{8}\Omega_{D}^{2}\Omega_{m_{0}}^{2}(1+z)^{6}\lambda^{2}\left\{-1+2M_{p}^{2}\Omega_{D}-M_{p}^{4}\Omega_{D}^{2}+9H_{0}^{4}M_{p}^{8}\Omega_{D}^{2}\Omega_{m_{0}}^{2}(1+z)^{6}\right\}
\end{equation*}
Here $\Omega_{m_0}=\frac{\rho_{m0}}{3 M_{p}^2 H_0^2}$ is the dimensionless fractional density parameter of matter, and $H_{0}$ is the present value of the Hubble parameter. Similarly $\Omega_{D}=\frac{\rho_{D}}{3 M_{p}^2 H_0^2}$ is the dimensionless fractional density parameter of NADE.

Similarly using eqns.(\ref{dgpfrw}), (\ref{nade}) and (\ref{denmat}) the FRW equations of DGP brane with NADE and CDM is obtained as,
\begin{equation}
H(z)=\frac{-\epsilon+\sqrt{\epsilon^{2}+4M_{p}^{2}H_{0}^{2}\Omega_{m_{0}}\kappa_{4}^{2}r_{c}^{2}(1+z)^{3}(1-\kappa_{4}^{2}M_{p}^{2}\Omega_{D})}}{2r_{c}(1-\kappa_{4}^{2}M_{p}^{2}\Omega_{D})}
\end{equation}


\subsection{Interacting Case}
Here we consider an interaction between the NADE and CDM. In this case, the continuity equations for matter and DE are respectively given by
\begin{equation}\label{IN1}
\dot{\rho}_{m}+3H\rho_{m}=\xi
\end{equation}
and
\begin{equation}
\dot{\rho}_{D}+3H\rho_{D}(1+w_{D})=-\xi
\end{equation}
Here, $\xi$ represents a variable that functions as the interaction component between CDM and NADE. A connection between DE and DM may serve as a potential solution to the cosmic coincidence issue \cite{int22, guo1}. The existence of this interaction can significantly alter the underlying dynamics and stability of the universe, resulting in an evolution that diverges from the Standard Cosmological Model. In reference \cite{int1}, Yang et al. examined extensive inhomogeneous perturbations and their corresponding instabilities within interacting DE models, revealing that the interaction term can profoundly influence the stability of cosmic perturbations. The phase-space dynamics of the FRW universe, encompassing both interacting and non-interacting fluids, were examined by Odintsov et al. in \cite{int2}, revealing that the interaction modifies the critical points and influences the late-time attractor characteristics. Zimdahl in \cite{int22} provided an extensive analysis of both linear and non-linear interactions within the dark sector, highlighting the theoretical justifications and phenomenological consequences, and asserting that interacting models can produce cosmological behaviors markedly distinct from the standard cosmological framework. Alternative research, including \cite{int3, int4, int5}, offers extensive theoretical foundations via modified gravity and holographic or agegraphic DE models, which frequently underpin the incorporation of interaction terms. These studies underscore that permitting interactions between DE and DM not only presents a viable solution to the coincidence problem but also enhances cosmic dynamics, perhaps elucidating some late-time events.

Generally, three frequently utilized variants of $\xi$ are extensively examined in research \cite{int1, int22, guo1}. They are given by
\begin{equation}
\xi_1 = 3 \delta H \rho_D,~ \quad \xi_2 = 3 \delta H (\rho_m + \rho_D),~ \quad \xi_3 = 3 \delta H \rho_m
\end{equation}
where $\delta$ denotes the interaction constant that signifies the intensity of the energy transfer between DM and DE. In this work, we utilize the third type of interaction, $\xi_3=3\delta H\rho_{m}$, which has been extensively employed in research to represent energy transfer within diverse cosmological limitations. The signature of the coupling constant $\delta$ dictates the trajectory of energy transfer. A positive value signifies that DE transitions into DM, while a negative value suggests that DM converts into DE. Observational evidence and theoretical analysis typically support the hypothesis that DE decays into DM. In \cite{guo1}, preliminary combined analysis (SNe + CMB shift + BAO) indicated that $\delta$ was minimal and frequently aligned with a value close to zero; for instance, in \cite{guo1} the authors documented $- 0.08 <\delta< 0.03$ for a constant coupling.

Solving equation (\ref{IN1}) using the interaction term $\xi_3$, we obtain
\begin{equation}\label{IN3}
\rho_{m}=\rho_{m_0}(1+z)^{3(1-\delta)}
\end{equation}
The corresponding FRW equation for the RS-II brane for the interacting case is obtained as,
\begin{equation}
H(z)=\pm\frac{1}{\sqrt{3}}\sqrt{\pm\frac{1}{\lambda^{2}\Omega_{D}^{2}}\left[\frac{\lambda}{M_{p}^{4}} (M_{p}^{2}\Omega_{D}-1)+3H_{0}^{2}\Omega_{D}\Omega_{m_{0}}(1+z)^{3-6\delta}+\frac{1}{2M_{p}^{4}}\lambda S\right]}
\end{equation}
where
\begin{equation*}
S=\sqrt{-12\lambda M_{p}^{4}\Omega_{D}^{2}\left(\frac{3H_{0}^{4}M_{p}^{4}\Omega_{m_{0}}^{2}(1+z)^{6-12\delta}}{\lambda}+2H_{0}^{2}M_{p}^{2}\Omega_{m_{0}}(1+z)^{3-6\delta}\right)+T^{2}}
\end{equation*}
and
\begin{equation*}
T=-2+2M_{p}^{2}\Omega_{D}+\frac{6H_{0}^{2}M_{p}^{4}\Omega_{D}\Omega_{m_{0}}(1+z)^{3-6\delta}}{\lambda}
\end{equation*}

Similarly the FRW equation for the DGP brane for the interacting case is given as,
\begin{equation}
H(z)=\frac{3\epsilon\pm3r_{c}\sqrt{\frac{\epsilon^{2}}{r_{c}^{2}}-4\kappa_{4}^{2}M_{p}^{2}H_{0}^{2}(-1+\kappa_{4}^{2}M_{p}^{2}\Omega_{D})\Omega_{m_{0}}(1+z)^{3-3\delta}}}{6(-1+\kappa_{4}^{2}M_{p}^{2}\Omega_{D})r_{c}}
\end{equation}

\section{Analysis of Cosmological Parameters}
\label{sec:params}
Now we will probe the cosmological aspects of NADE within the braneworld framework that we have already established in the previous section. To accomplish this, we shall examine fundamental cosmological parameters such as the equation of state parameter, and the deceleration parameter for both RS-II and DGP brane models.

\subsection{EoS parameter}
The equation of state (EoS) parameter is a critical quantity for delineating the dynamic characteristics of the DE component and is expressed as $w_D=\frac{p_D}{\rho_D}$. Its numerical value offers a clear idea of the cosmic dynamics of the DE component. Specifically, $w_D=-1$ signifies the behavior of a cosmological constant, while $w_D>-1$ and $w_D<-1$ denote the quintessence and phantom phases, respectively \cite{cc2, de2}. The evolution of $w_D$ is hence beneficial for assessing if the NADE component demonstrates dynamic characteristics and if it can transition across the phantom division $w_D=-1$. Within the current braneworld paradigm, the altered gravitational dynamics can affect the evolution of $w_D$ via additional higher-dimensional adjustments, thereby enabling the exploration of the impact of the extra dimension on DE behavior. We will study the EoS parameter for both the non-interacting and interacting scenarios.

\subsubsection{Non-interacting case}
For this case the EoS parameter for both RS-II brane and DGP brane is obtained as,
\begin{equation}
w_{D}=-1+\frac{2\sqrt{\Omega_{D}}(1+z)}{n}
\end{equation}
The EoS parameter of non-interacting NADE adopts the same functional form in both the RS-II and DGP braneworld scenarios, which is an intriguing aspect of the current investigation. Specifically, the explicit dependence of $w_D$ on braneworld parameters, such as the brane tension $\lambda$, crossover scale $r_c$, and branch parameter $\epsilon$ vanishes, leaving $w_D$ directly controlled by the fractional NADE density $\Omega_D$ and the agegraphic parameter $n$. This suggests that rather than the specific modification of the Friedmann equation, the inherent agegraphic prescription $\rho_{D} \propto \eta^{-2}$ is the primary determinant of the instantaneous EoS \cite{nd1}. However, this does not imply total insensitivity to braneworld gravity in the absence of an explicit reliance. The dynamical evolution of $H$ and $\Omega_D$ allow the braneworld parameters to enter implicitly, which can alter the redshift evolution of $w_D$.

We notice that $\frac{2\sqrt{\Omega_{D}}(1+z)}{n}>0$ for $n>0,~ 0<\Omega_{D}<1, ~z>-1$. Therefore we have $w_{D}>-1$ throughout the physical domain. Thus, non-interacting NADE behaves as a quintessence-like DE component and does not cross the phantom divide $w_{D}=-1$ in this formulation. Here, the parameter $n$ is the primary direct parameter controlling the deviation from a cosmological-constant-like behavior.
From the expression we see that $w_{D}+1 \propto 1/n$. Therefore, increasing $n$ suppresses the departure from $w_D=-1$. Conversely for smaller values of $n$, we have large deviation from $w_D=-1$. Thus we see that the NADE parameter $n$ controls the strength of the dynamical nature of NADE. The expression also shows another interesting qualitative picture. At late times $z\to -1$, so $w_{D}\to -1$, provided $\Omega_{D}$ remains finite. Thus NADE naturally approaches a cosmological-constant-like state in the far future. At earlier times, $(1+z)$ is larger and $\Omega_D$ is generally smaller. The competition between these two factors determines the actual evolution of $w_D$. Despite having the same analytical EoS relation, RS-II and DGP branes can create distinct histories of $\Omega_{D}(z)$, and consequently distinct histories of $w_{D}(z)$.

\begin{figure}
\centering
\includegraphics[height=2.5in,width=3.5in]{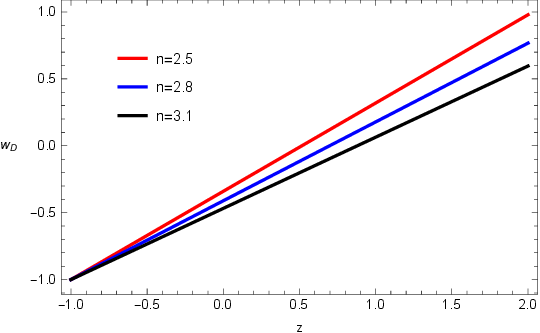}

\vspace{1 mm}

\caption{Graphical portrayal of $w_{D}$ versus $z$ for non-interacting case with $\Omega_{D}=0.68$ and the best fit parameter values from tables \ref{tab:rsii_nonint} and \ref{tab:constraints_DGP_nonint} for both RS-II and DGP brane.}\label{F1}
\vspace{2 mm}

\end{figure}

To understand the behavior, we have generated the plot of the EoS parameter in Fig.\ref{F1} using the best fit parameter values from tables \ref{tab:rsii_nonint} and \ref{tab:constraints_DGP_nonint}. It shows three different curves for different value of the NADE parameter $n$. We observe the transition of the radiation era at high redshift (early universe) to the quintessence behavior at low redshift (late universe) crossing the dust era in between. So, this model concurs with the observed evolution of the universe. Moreover, as the value of $n$ increases, the value of EoS parameter decreases. But this dependence gradually vanishes as we move to smaller redshifts.

\subsubsection{Interacting case}
For the interacting case, the EoS for the RS-II brane model is obtained as
\begin{equation}
w_{D}=-1+\frac{2\sqrt{\Omega_{D}}(1+z)}{3n}-\delta\left(1\pm\frac{3H_{0}^{2}\lambda^{2}\Omega_{D}^{3}\Omega_{m_{0}}(1+z)}{-\frac{\lambda}{M_{p}^{4}}+\frac{\lambda\Omega_{D}}{M_{p}^{2}}+\frac{\lambda S}{2M_{p}^{4}}+3H_{0}^{2}\Omega_{D}\Omega_{m_{0}}(1+z)^{3-6\delta}}\right).
\end{equation}

\begin{figure}
\centering
      \includegraphics[height=2.5in,width=3.5in]{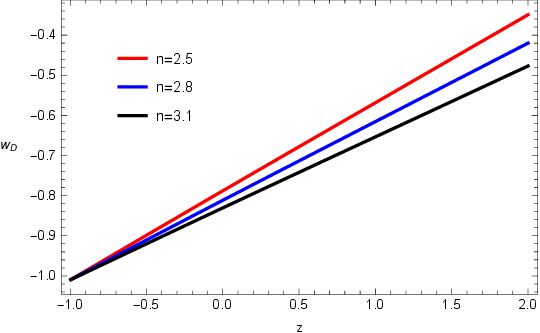}

      \vspace{1 mm}

      \caption{Graphical portrayal of $w_{D}$ versus $z$ for interacting NADE in RS-II brane with $\Omega_{D}=0.68,~\Omega_{m_{0}}=0.28,~\delta=0.01,~\lambda=1.4585,~H_{0}=69.25$.}\label{F2}
      \vspace{2 mm}

      \end{figure}

The above has been plotted in figure \ref{F2} for the best fit parameter values given in table \ref{tab:rsii_int}. It shows three different curves as the NADE parameter $n$ is varied. We observe the transition of dust era at high redshift to quintessence behavior at low redshift. So, this model concurs with the observations. We also see that the EoS parameter values varies inversely to the values of $n$. 

The Eos parameter for DGP brane model for the interacting case is obtained as
\begin{equation}
w_{D}=-1+\frac{2\sqrt{\Omega_{D}}(1+z)}{3n}-\delta\left(1+\frac{4H_{0}^{2}\Omega_{D}(-1+\kappa_{4}^{2}M_{p}^{2}\Omega_{D})^{2}\Omega_{m_{0}}r_{c}^{2}(1+z)}{\left(\epsilon\pm r_{c}\sqrt{\frac{\epsilon^{2}}{r_{c}^{2}}-4H_{0}^{2}\kappa_{4}^{2}M_{p}^{2}(-1+\kappa_{4}^{2}M_{p}^{2}\Omega_{D})\Omega_{m_{0}}(1+z)^{3-3\delta}}\right)^{2}}\right).
\end{equation}

\begin{figure}
      \centering
      \includegraphics[height=2.5in,width=3.5in]{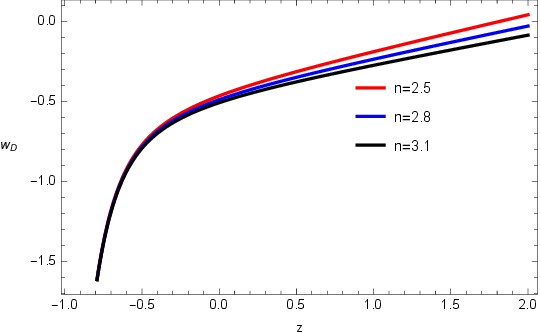}

      \vspace{1 mm}

      \caption{Graphical portrayal of $w_{D}$ versus $z$ for interacting NADE in DGP brane with $\Omega_{D}=0.68,~\Omega_{m_{0}}=0.28,~\epsilon=1,~\kappa_{4}=5,~H_{0}=69.25,~\delta=0.01$.}\label{F3}
      \vspace{2 mm}

      \end{figure}

This has been plotted in figure \ref{F3} using the best fit parameter values from table \ref{tab:constraints}. It shows three different curves as the NADE parameter $n$ is varied. We observe the transition to quintessence and finally to the phantom phase at low redshift from the dust era at high redshift crossing the phantom divide line. So the model is in agreement with the observations. We also see that the EoS parameter values varies inversely to the values of $n$.

\subsection{Deceleration Parameter}
The deceleration parameter provides a complementary description of the expansion history of the Universe and is defined by
\begin{equation}
q=-\frac{\ddot{a}}{aH^{2}}=-1-\frac{\dot{H}}{H^{2}}
\end{equation}
The value of $q$ distinguishes between a decelerated and an accelerated expansion of the universe. $q>0$ signifies a decelerating Universe, whereas $q<0$ denotes an accelerated expansion, and $q=0$ shows the transition between these two states. Consequently, the evolution of $q$ offers an effective means to discern the shift from slowdown to acceleration in the history of cosmic expansion \cite{acc1, acc3}. In the current NADE-braneworld framework, the altered Friedmann dynamics and the progression of the agegraphic DE density dictate the characteristics of $q$. Analyzing $q$ in relation to redshift or the scale factor enables us to investigate the influence of higher-dimensional modifications on the history of expansion and the initiation of late-time cosmic acceleration. Here also we study the non-interacting and interacting cases separately.

\subsubsection{Non-interacting case}
For the RS-II brane, the expression for the deceleration parameter is obtained as
\begin{equation}
q=-1+\frac{M_{p}^{2}(1+z)}{2nH^{2}}\left(3H_{0}^{2}\Omega_{m_{0}}(1+z)^{2}n+2H^{2}\Omega_{D}^{3/2}\right)\left[1+\frac{3M_{p}^{2}}{\lambda}\left(H_{0}^{2}\Omega_{m_{0}}(1+z)^{3}+H^{2}\Omega_{D}\right)\right]
\end{equation}

\begin{figure}
      \centering
      \includegraphics[height=2.5in,width=3.5in]{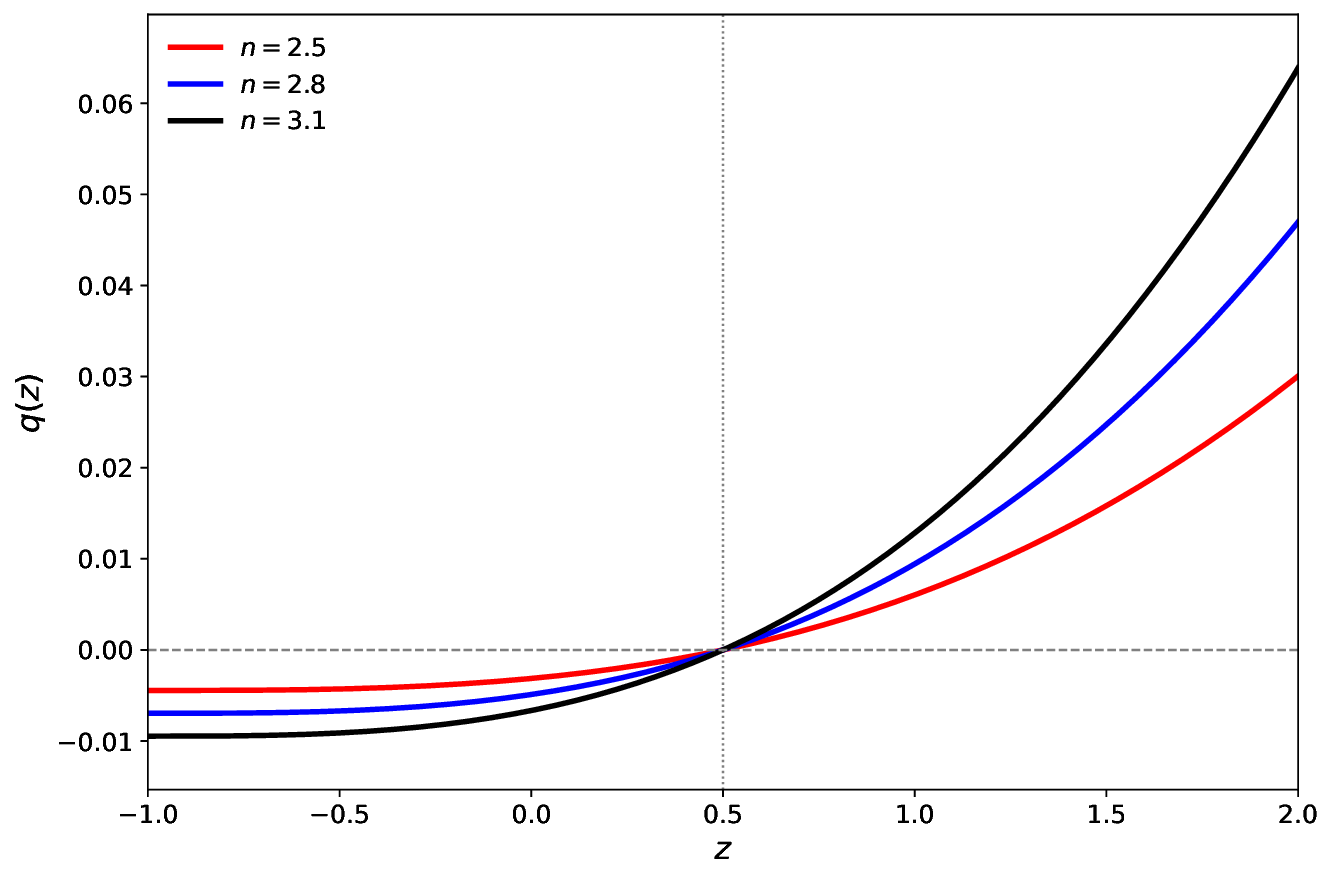}

      \vspace{1 mm}

      \caption{Graphical portrayal of $q$ versus $z$ for non-interacting NADE in RS-II brane with $\Omega_{D}=0.68,~\Omega_{m_{0}}=0.28,~\lambda=1.4585,~H_{0}=69.25$.}\label{F4}
      \vspace{2 mm}

      \end{figure}

Similarly, for the DGP brane model the expression for $q$ is obtained as,
\begin{equation}
q=-1+\frac{\kappa_{4}^{2}M_{p}^{2}(1+z)}{nH(2H+\frac{\epsilon}{r_{c}})}\Big\{3nH_{0}^{2}\Omega_{m_{0}}(1+z)^{2}+2H^{2}\Omega_{D}^{3/2}\Big\}
\end{equation}

\begin{figure}
      \centering
      \includegraphics[height=2.5in,width=3.5in]{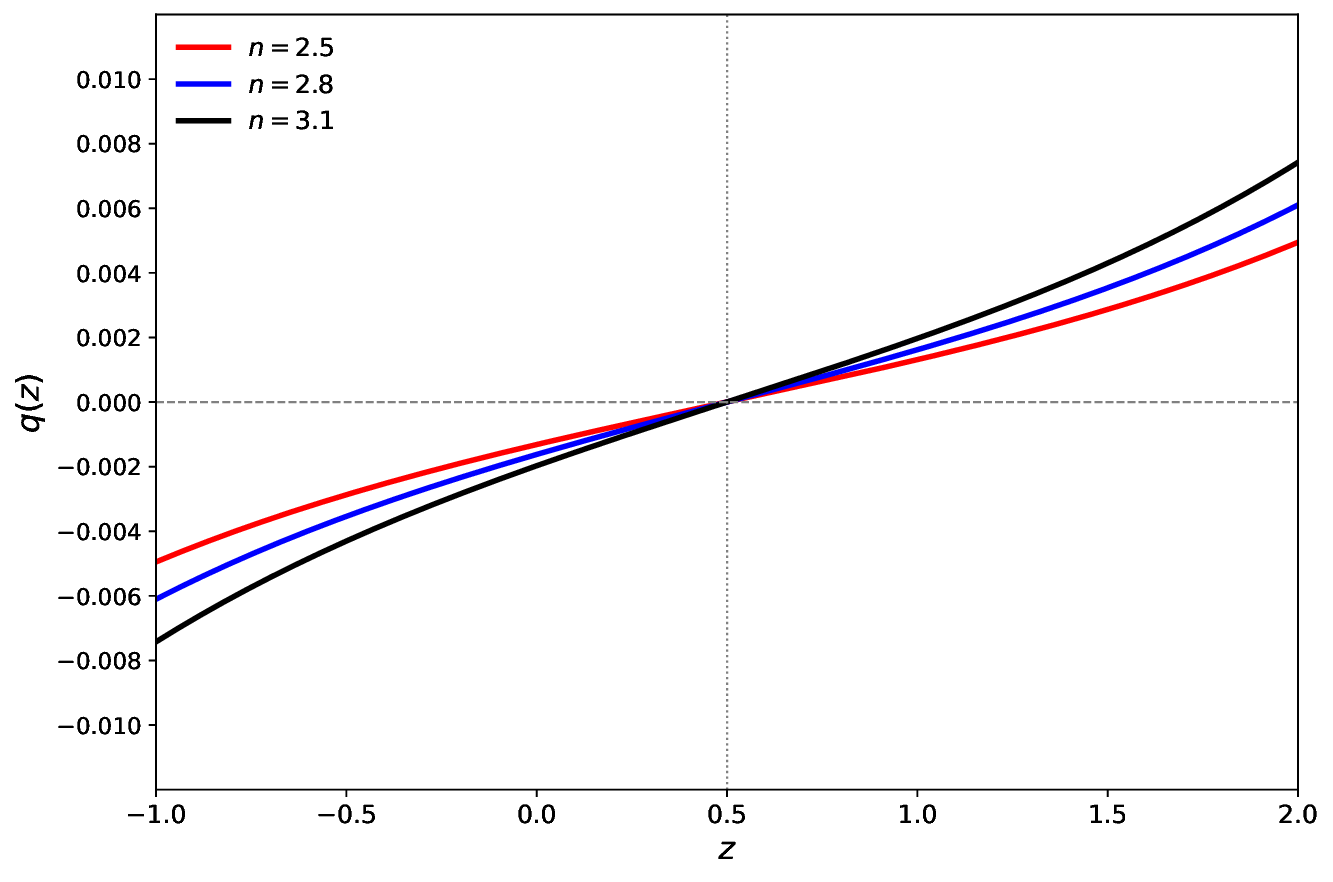}

      \vspace{1 mm}

      \caption{Graphical portrayal of $q$ versus $z$ for non-interacting NADE in DGP brane with $\Omega_{D}=0.68,~\Omega_{m_{0}}=0.28,~\epsilon=1,~\kappa_{4}=5,~H_{0}=69.25$.}\label{F5}
      \vspace{2 mm}

\end{figure}

The plots of the decelaration parameter are given in Figures \ref{F4} and \ref{F5} for the non-interacting NADE in RS-II and DGP branes respectively using the best fit parameter values in Tables \ref{tab:rsii_nonint} and \ref{tab:constraints_DGP_nonint}. Both plot show three different curves as the NADE parameter $n$ is varied. We observe the transition from early universe decelerated to the late universe accelerated scenario at $z\approx 0.5$. These two profiles correspond to the observational data \cite{rs15, rs16}.

\subsubsection{Interacting case}
The expression for the deceleration parameter for the RS-II brane for the interacting case is given by
\begin{eqnarray*}
q=-1-\frac{3M_{p}^{2}(1+z)}{nH^{2}}\left(3(1-\delta)H_{0}^{2}\Omega_{m_{0}}(1+z)^{2-3\delta}n-2H^{2}\Omega_{D}^{3/2}\right)\Big[1
\end{eqnarray*}
\begin{equation}
+\frac{3M_{p}^{2}}{\lambda}\left(H_{0}^{2}\Omega_{m_{0}}(1+z)^{3(1-\delta)}+H^{2}\Omega_{D}\right)\Big]
\end{equation}

\begin{figure}
      \centering
      \includegraphics[height=2.5in,width=3.5in]{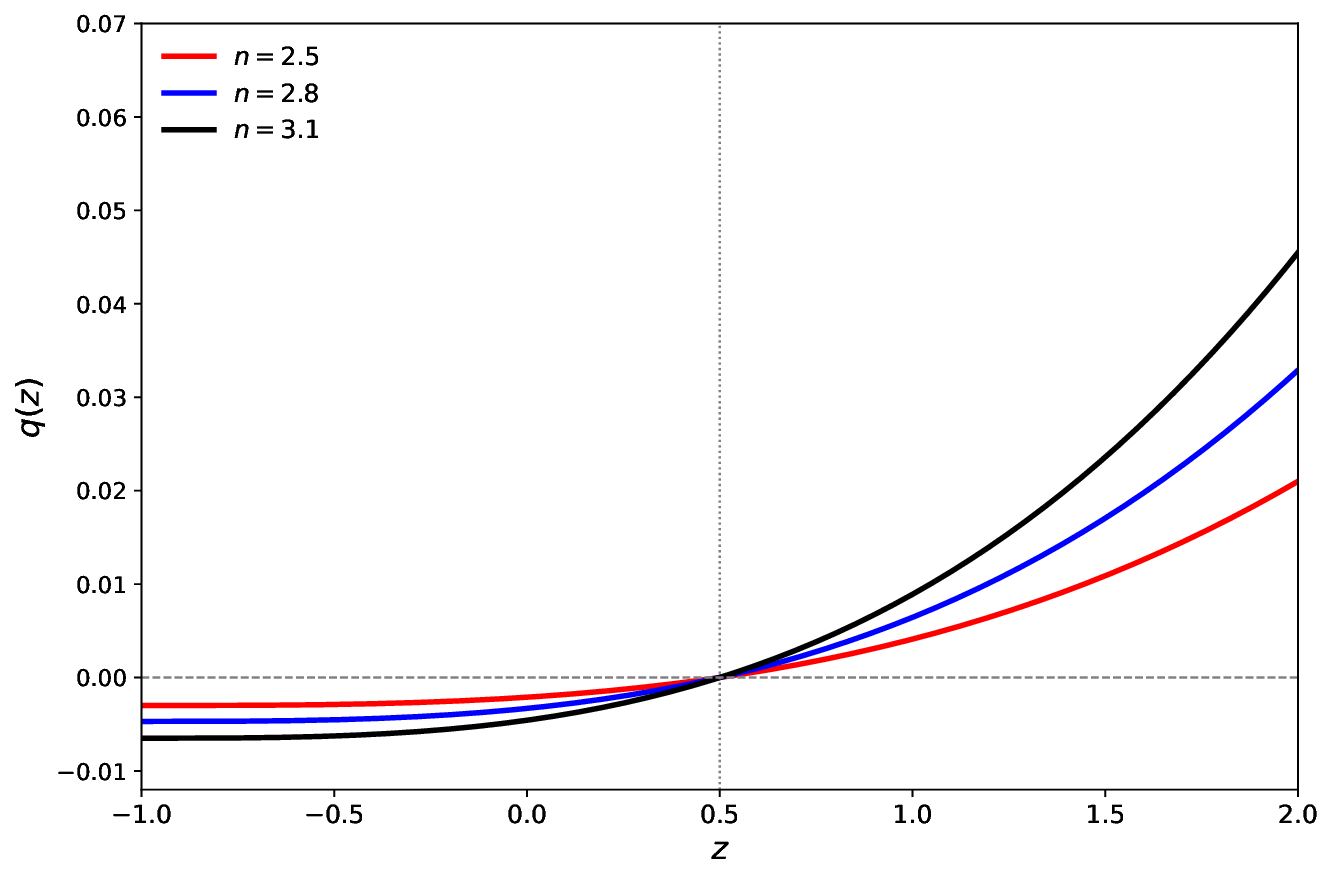}

      \vspace{1 mm}

      \caption{Graphical portrayal of $q$ versus $z$ for interacting NADE in RS-II brane with $\Omega_{D}=0.68,~\Omega_{m_{0}}=0.28,~\delta=0.01,~\lambda=1.4585,~H_{0}=69.25$.}\label{F6}
      \vspace{2 mm}

      \end{figure}
    
The corresponding expression for the DGP brane is given by
\begin{equation}
q=-1+\frac{\kappa_{4}^{2}M_{p}^{2}(1+z)}{nH^{2}(2H+\frac{\epsilon}{r_{c}})}\Big\{3n(1-\delta)H_{0}^{2}\Omega_{m_{0}}(1+z)^{2-3\delta}+2H^{3}\Omega_{D}^{3/2}\Big\}
\end{equation}

    \begin{figure}
      \centering
      \includegraphics[height=2.5in,width=3.5in]{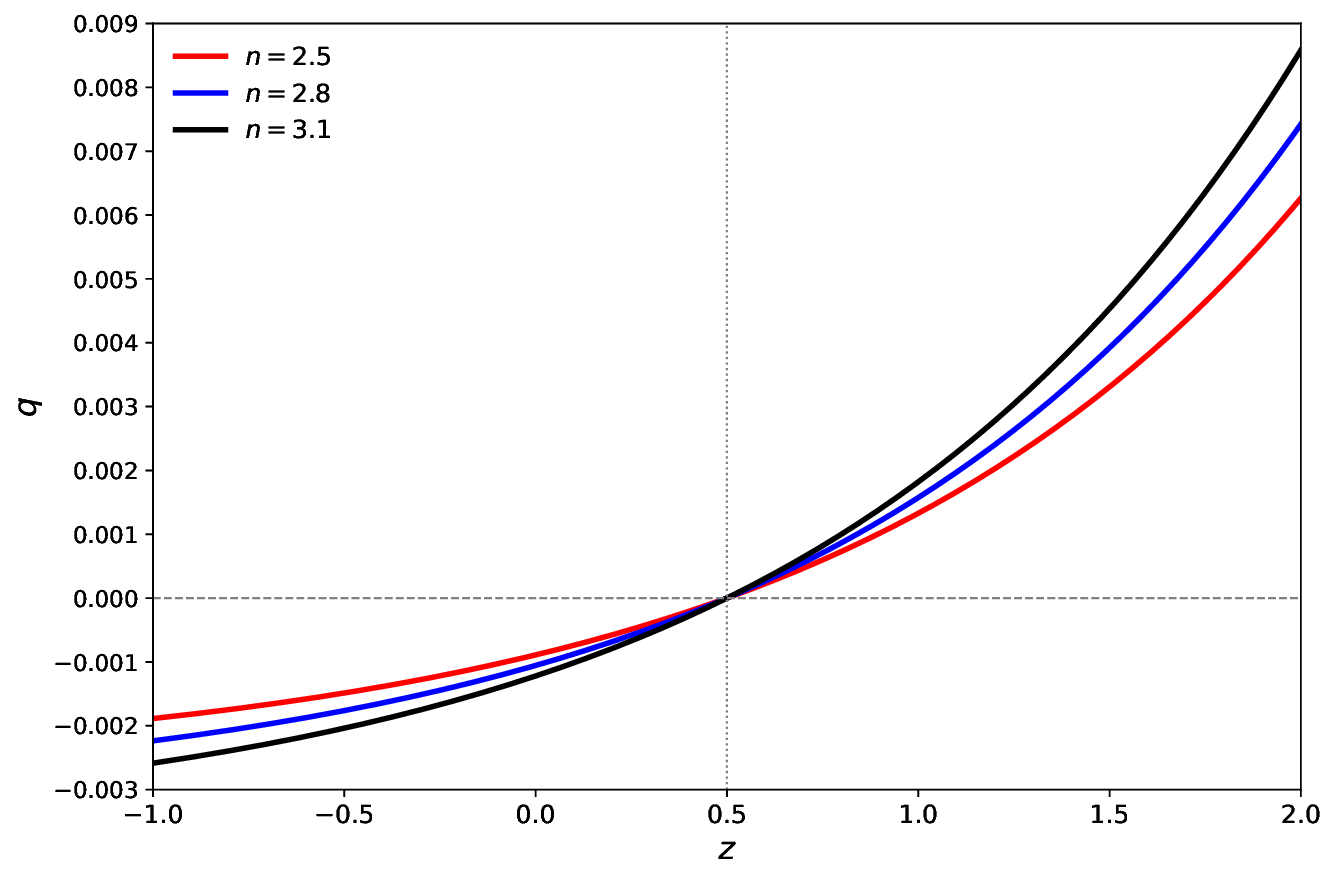}

      \vspace{1 mm}

      \caption{Graphical portrayal of $q$ versus $z$ for interacting NADE in DGP brane with $\Omega_{D}=0.68,~\Omega_{m_{0}}=0.28,~\epsilon=1,~\kappa_{4}=5,~H_{0}=69.25,~\delta=0.01$.}\label{G1}
      \vspace{2 mm}

      \end{figure}

The plots for the decelaration parameter are given in Figures \ref{F6} and \ref{G1} for the interacting NADE in RS-II and DGP branes respectively. Here we have used best fit parameter values from the tables \ref{tab:rsii_int} and \ref{tab:constraints}. The plots show three different curves as the NADE parameter $n$ is varied. We observe the transition from the early universe decelerated behavior to late universe accelerated behavior at $z\approx 0.5$, which correspond to the observations \cite{rs15, rs16}.

\subsection{Stability Analysis}
We now investigate the stability of the developed system using the squared speed of sound, characterized as the partial derivative of pressure about energy density \cite{nda4, nda5}. The expression for the square of the speed of sound is given as follows
\begin{equation}
v_{s}^{2}=\frac{\partial p}{\partial\rho}
\end{equation}
where $\rho=\rho_{m}+\rho_{D}$ is the total energy density and $p=p_{m}+p_{D}$ is the total pressure. For CDM, $p_{m}=0$, so, $p$ becomes $p_{D}$ only. With this, the above equation reduces to
\begin{equation}
v_{s}^{2}=\frac{\partial p_{D}}{\partial (\rho_{m}+\rho_{D})}=\frac{\dot{p}_{D}}{\dot{\rho}_{m}+\dot{\rho}_{D}}
\end{equation}
where dot represents derivative with respect to time.

\subsubsection{Non-interacting case}
The expression for $v_{s}^2$ for the non-interacting case can be given by,
\begin{equation}\label{vsn}
v_{s}^{2}=\frac{H^{2}\Omega_{D}\{(1+z)\frac{dw_{D}}{dz}-3w_{D}(1+w_{D})\}}{3\{H_{0}^{2}\Omega_{m_{0}}(1+z)^{3}+(1+w_{D})H^{2}\Omega_{D}\}}
\end{equation}

Using the above expression, the squared speed of sound for the RS-II brane is given by
\begin{eqnarray*}
v_{s}^{2}=\pm\left[\left\{4\left(2n-3\sqrt{\Omega_{D}}(1+z)\right)\left(-\lambda+\lambda M_{p}^{2}\Omega_{D}+3H_{0}^{2}M_{p}^{4}\Omega_{D}\Omega_{m_{0}}(1+z)^{3}+H_{0}M_{p}^{2}\sqrt{\Pi-\lambda^{2}\Sigma}\right)\right\}\right]/
\end{eqnarray*}
\begin{equation}\label{vsn_RSII}
\left[3n\left\{2\lambda-2\lambda M_{p}^{2}\Omega_{D}+3H_{0}^{2}\lambda^{2}M_{p}^{4}n\sqrt{\Omega_{D}}\Omega_{m_{0}}(1+z)^{2}+6H_{0}^{2}M_{p}^{4}\Omega_{D}\Omega_{m_{0}}(1\pm z)^{3}-2H_{0}M_{p}^{2}\sqrt{\Pi+\lambda^{2}\Sigma}\right\}\right]
\end{equation}

where $\Pi=-6\lambda^{3}M_{p}^{2}\Omega_{D}^{2}\Omega_{m_{0}}(1+z)^{3}+6\lambda\Omega_{D}(-1+M_{p}^{2}\Omega_{D})\Omega_{m_{0}}(1+z)^{3}+9H_{0}^{2}M_{p}^{4}\Omega_{D}^{2}\Omega_{m_{0}}^{2}(1+z)^{6}$ and $\Sigma=-1+2M_{p}^{2}\Omega_{D}-M_{p}^{4}\Omega_{D}^{2}+9H_{0}^{4}M_{p}^{8}\Omega_{D}^{2}\Omega_{m_{0}}^{2}(1+z)^{6}$.

Similarly the squared speed of sound for the DGP brane is given by
\begin{eqnarray*}
v_{s}^{2}=\left[\Omega_{D}^{3/2}(1+z)\left(2n-3\sqrt{\Omega_{D}}(1+z)\right)\left(3\epsilon\mp\sqrt{\epsilon^{2}-4H_{0}^{2}\kappa_{4}^{2}M_{p}^{2}(-1+\kappa_{4}^{2}M_{p}^{2}\Omega_{D})\Omega_{m_{0}}r_{c}^{2}(1+z)^{3}}\right)^{2}\right]/    
\end{eqnarray*}
\begin{eqnarray*}
\Bigg[27n^{2}r_{c}^{2}(1-\kappa_{4}^{2}M_{p}^{2}\Omega_{D})^{2}\Bigg\{H_{0}^{2}\Omega_{m_{0}}(1+z)^{3}
\end{eqnarray*}
    
\begin{equation}\label{vsn_DGP}
+\frac{\Omega_{D}^{3/2}(1+z)\left(3\epsilon\mp\sqrt{3}\sqrt{\epsilon^{2}-4H_{0}^{2}\kappa_{4}^{2}M_{p}^{2}(-1+\kappa_{4}^{2}M_{p}^{2}\Omega_{D})\Omega_{m_{0}}r_{c}^{2}(1+z)^{3}}\right)^{2}}{18n(-1+\kappa_{4}^{2}M_{p}^{2}\Omega_{D})^{2}r_{c}^{2}}\Bigg\}\Bigg]
\end{equation}

\begin{figure}
      \centering
      \includegraphics[height=2.5in,width=3.5in]{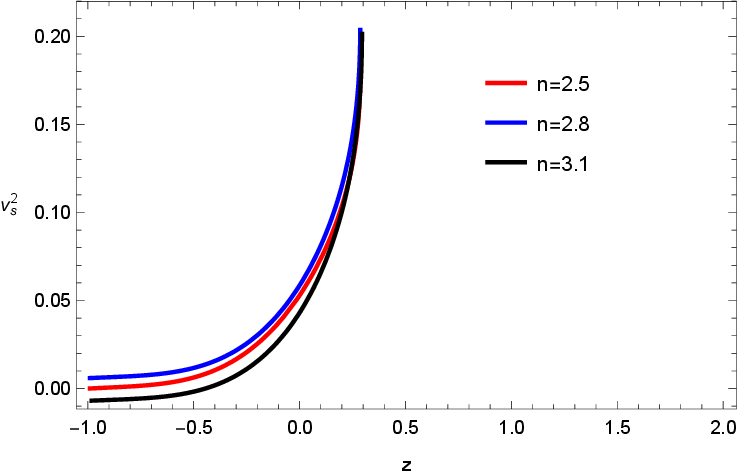}

      \vspace{1 mm}

      \caption{Graphical portrayal of $v_{s}^{2}$ versus $z$ for non-interacting NADE in RS-II brane with $\Omega_{D}=0.68,~\Omega_{m_{0}}=0.28,~\lambda=1.4585,~H_{0}=69.25$.}\label{F7}
      \vspace{2 mm}

      \end{figure}

      \begin{figure}
      \centering
      \includegraphics[height=2.5in,width=3.5in]{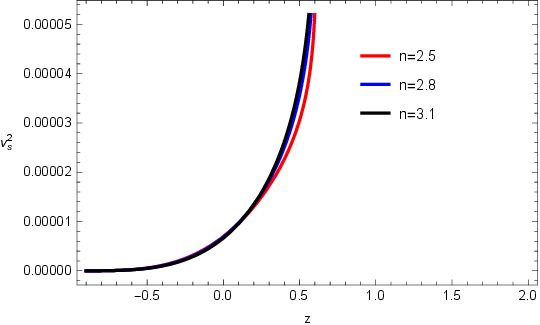}

      \vspace{1 mm}

      \caption{Graphical portrayal of $v_{s}^{2}$ versus $z$ for non-interacting NADE in DGP brane with $\Omega_{D}=0.68,~\Omega_{m_{0}}=0.28,~\epsilon=1,~\kappa_{4}=5,~H_{0}=69.25$..}\label{F8}
      \vspace{2 mm}

      \end{figure}

The plots of the squared sound speed are provided in Figures \ref{F7} and \ref{F8} for the non-interacting NADE in RS-II and DGP branes respectively. The best fit values from the tables \ref{tab:rsii_nonint} and \ref{tab:constraints_DGP_nonint} have been used. The plots have been generated for different values of the NADE parameter $n$. We observe that $v_{s}^{2}>0$ for all values of $z$. These plots imply a classically stable scenario throughout the cosmological evolution.

\subsubsection{Interacting case}
The expression for $v_{s}^2$ for the interacting case can be given by,
\begin{equation}\label{vsi}
v_{s}^{2}=\frac{H^{2}\Omega_{D}\{(1+z)\frac{dw_{D}}{dz}-3w_{D}(1+w_{D})\}}{3\{H_{0}^{2}\Omega_{m_{0}}(1+z)^{3(1-\delta)}+(1+w_{D})H^{2}\Omega_{D}\}}
\end{equation}
Using the above expression, the squared speed of sound for the RS-II brane is given by
\begin{eqnarray*}
v_{s}^{2}=\pm\Bigg[\left(-\frac{\lambda}{M_{p}^{4}}+\frac{\lambda\Omega_{D}}{M_{p}^{2}}+\frac{\lambda S}{2M_{p}^{4}}+3H_{0}^{2}\Omega_{D}\Omega_{m_{0}}(1+z)^{3-6\delta}\right)\left\{(1+z)\left(\frac{2\sqrt{\Omega_{D}}}{3n}-\left(3\delta H_{0}^{2}\lambda^{2}\Omega_{D}^{3}\Omega_{m_{0}}\right.\right.\right.
\end{eqnarray*}
\begin{eqnarray*}
\left.\left.\left.\left(-\frac{\lambda}{M_{p}^{4}}+\frac{\lambda\Omega_{D}}{M_{p}^{2}}+\frac{\lambda S}{2M_{p}^{4}}-6H_{0}^{2}\Omega_{D}\Omega_{m_{0}}(1+z)^{3-6\delta}(1\pm3\delta)\right)\right)\Big/\left(-\frac{\lambda}{M_{p}^{4}}+\frac{\lambda\Omega_{D}}{M_{p}^{2}}+\frac{\lambda S}{2M_{p}^{4}}\right.\right.\right.
\end{eqnarray*}
\begin{eqnarray*}
\left.\left.\left.+3H_{0}^{2}\Omega_{D}\Omega_{m_{0}}(1+z)^{3-6\delta}\right)^{2}\right)-3\left\{-1+\frac{2\sqrt{\Omega_{D}}(1+z)}{3n}-\delta\left(1\right.\right.\right.
\end{eqnarray*}
\begin{eqnarray*}
\left.\left.\left.+\frac{3H_{0}^{2}\lambda^{2}\Omega_{D}^{3}\Omega_{m_{0}}(1+z)}{-\frac{\lambda}{M_{p}^{4}}+\frac{\lambda\Omega_{D}}{M_{p}^{2}}+\frac{\lambda S}{2M_{p}^{4}}+3H_{0}^{2}\Omega_{D}\Omega_{m_{0}}(1+z)^{3-6\delta}}\right)\right\}\left\{\frac{2\sqrt{\Omega_{D}}(1+z)}{3n}-\delta\left(1\right.\right.\right.
\end{eqnarray*}
\begin{eqnarray*}
\left.\left.\left.+\frac{3H_{0}^{2}\lambda^{2}\Omega_{D}^{3}\Omega_{m_{0}}(1+z)}{-\frac{\lambda}{M_{p}^{4}}+\frac{\lambda\Omega_{D}}{M_{p}^{2}}+\frac{\lambda S}{2M_{p}^{4}}+3H_{0}^{2}\Omega_{D}\Omega_{m_{0}}(1+z)^{3-6\delta}}\right)\right\}\right\}\Bigg]\Bigg/\Bigg[9\lambda^{2}\Omega_{D}\left\{H_{0}^{2}\Omega_{m_{0}}(1+z)^{3-3\delta}\right.
\end{eqnarray*}
\begin{eqnarray*}
\left.+\frac{1}{3\lambda^{2}\Omega_{D}}\left(-\frac{\lambda}{M_{p}^{4}}+\frac{\lambda\Omega_{D}}{M_{p}^{2}}+\frac{\lambda S}{2M_{p}^{4}}+3H_{0}^{2}\Omega_{D}\Omega_{m_{0}}(1+z)^{3-6\delta}\right)\left\{\frac{2\sqrt{\Omega_{D}}(1+z)}{3n}-\delta\left(1\right.\right.\right.
\end{eqnarray*}
\begin{equation}\label{vsi_RSII}
\left.\left.\left.+\frac{3H_{0}^{2}\lambda^{2}\Omega_{D}^{3}\Omega_{m_{0}}(1+z)}{-\frac{\lambda}{M_{p}^{4}}+\frac{\lambda\Omega_{D}}{M_{p}^{2}}+\frac{\lambda S}{2M_{p}^{4}}+3H_{0}^{2}\Omega_{D}\Omega_{m_{0}}(1+z)^{3-6\delta}}\right)\right\}\right\}\Bigg]
\end{equation}

\begin{figure}
      \centering
      \includegraphics[height=2.5in,width=3.5in]{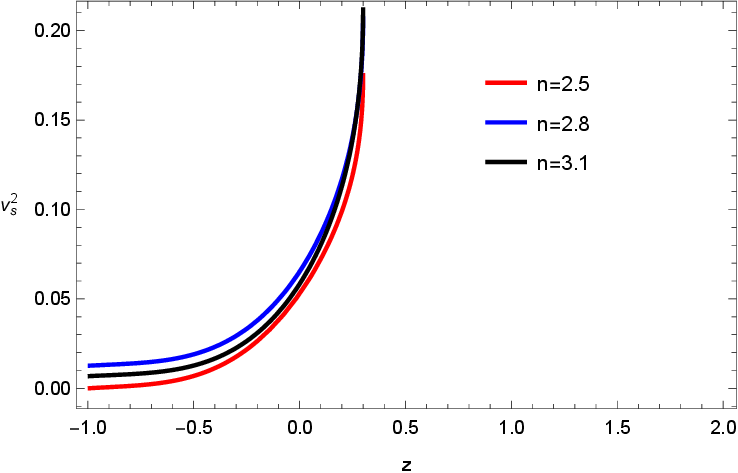}

      \vspace{1 mm}

      \caption{Graphical portrayal of $v_{s}^{2}$ versus $z$ for interacting NADE in RS-II brane with $\Omega_{D}=0.68,~\Omega_{m_{0}}=0.28,~\delta=0.01,~\lambda=1.4585,~H_{0}=69.25$.}\label{F9}
      \vspace{2 mm}

      \end{figure}

Similarly, the squared speed of sound for the DGP brane is given by
\begin{eqnarray*}
v_{s}^{2}=\Bigg[\Omega_{D}\Bigg\{3\epsilon+\sqrt{3}\sqrt{3\epsilon^{2}+4H_{0}^{2}\kappa_{4}^{2}M_{p}^{2}(-1+\kappa_{4}^{2}M_{p}^{2}\omega_{D})\omega_{m_{0}}r_{c}^{2}(1+z)^{3-3\delta}}\Bigg\}^{2}\Bigg\{(1+z)\left(\frac{2\sqrt{\omega_{D}}}{3n}\right.
\end{eqnarray*}  
\begin{eqnarray*}
\left.-\left(36\delta H_{0}^{2}\Omega_{D}(-1+\kappa_{4}^{2}M_{p}^{2}\Omega_{D})^{2}\Omega_{m_{0}}r_{c}^{2}\left(3\epsilon+r_{c}\left(12\sqrt{3}(-1+\delta)H_{0}^{2}\kappa_{4}^{2}M_{p}^{2}(-1+\kappa_{4}^{2}M_{p}^{2}\omega_{D})\right.\right.\right.\right.   
\end{eqnarray*}
\begin{eqnarray*}
\left.\left.\left.\left.\Omega_{m_{0}}r_{c}(1+z)^{3-3\delta}\Big/\sqrt{3\epsilon^{2}+4H_{0}^{2}\kappa_{4}^{2}M_{p}^{2}(-1+\kappa_{4}^{2}M_{p}^{2}\omega_{D})\Omega_{m_{0}}r_{c}^{2}(1+z)^{3-3\delta}}+\sqrt{3}\right.\right.\right.\right.   
\end{eqnarray*}
\begin{eqnarray*}
\left.\left.\left.\left.\sqrt{3\epsilon^{2}+4H_{0}^{2}\kappa_{4}^{2}M_{p}^{2}(-1+\kappa_{4}^{2}M_{p}^{2}\Omega_{D})\Omega_{m_{0}}r_{c}^{2}(1+z)^{3-3\delta}}\right)\right)\Bigg/\left(3\epsilon\right.\right.\right.    
\end{eqnarray*}
\begin{eqnarray*}
\left.\left.\left.+3\sqrt{3}\sqrt{3\epsilon^{2}+4H_{0}^{2}\kappa_{4}^{2}M_{p}^{2}(-1+\kappa_{4}^{2}M_{p}^{2}\omega_{D})\Omega_{m_{0}}}\left(-1+\frac{2\sqrt{\Omega_{D}}(1+z)}{3n}-\delta\left(1\right.\right.\right.\right.\right.   
\end{eqnarray*}
\begin{eqnarray*}
\left.\left.\left.\left.\left.+\frac{36H_{0}^{2}\Omega_{D}(-1+\kappa_{4}^{2}M_{p}^{2}\omega_{D})^{2}\Omega_{m_{0}}r_{c}^{2}(1+z)}{\left(3\epsilon+\sqrt{3}\sqrt{3\epsilon^{2}+4H_{0}^{2}\kappa_{4}^{2}M_{p}^{2}(-1+\kappa_{4}^{2}M_{p}^{2}\Omega_{D})\Omega_{m_{0}}r_{c}^{2}(1+z)^{3-3\delta}}\right)^{2}}\right)\right)\left(\frac{2\sqrt{\Omega_{D}}(1+z)}{3n}-\delta\left(1\right.\right.\right.\right.\right.
\end{eqnarray*}
\begin{eqnarray*}
\left.\left.\left.\left.\left.+\frac{36H_{0}^{2}\Omega_{D}(-1+\kappa_{4}^{2}M_{p}^{2}\omega_{D})^{2}\Omega_{m_{0}}r_{c}^{2}(1+z)}{\left(3\epsilon+\sqrt{3}\sqrt{3\epsilon^{2}+4H_{0}^{2}\kappa_{4}^{2}M_{p}^{2}(-1+\kappa_{4}^{2}M_{p}^{2}\Omega_{D})\Omega_{m_{0}}r_{c}^{2}(1+z)^{3-3\delta}}\right)^{2}}\right)\right)\right)\right)\Bigg/\right.
\end{eqnarray*}
\begin{eqnarray*}
\left.\left(108(-1+\kappa_{4}^{2}M_{p}^{2}\Omega_{D})^{2}r_{c}^{2}\left(H_{0}^{2}\Omega_{m_{0}}(1+z)^{3-3\delta}+\Bigg(\Omega_{D}\Big(3\epsilon^{2}+4H_{0}^{2}\kappa_{4}^{2}M_{p}^{2}(-1+\kappa_{4}^{2}M_{p}^{2}\Omega_{D})\Omega_{m_{0}}r_{c}^{2}\right.\right.\right.
\end{eqnarray*}
\begin{eqnarray*}
\left.\left.\left.(1+z)^{3-3\delta}\frac{36H_{0}^{2}\Omega_{D}(-1+\kappa_{4}^{2}M_{p}^{2}\Omega_{D})^{2}\Omega_{m_{0}}r_{c}^{2}(1+z)}{\left(3\epsilon^{2}+4H_{0}^{2}\kappa_{4}^{2}M_{p}^{2}(-1+\kappa_{4}^{2}M_{p}^{2}\Omega_{D})\Omega_{m_{0}}r_{c}^{2}(1+z)^{3-3\delta}\right)^{2}}\Big)\Bigg)\right)\right.\right.
\end{eqnarray*}
\begin{equation}\label{vsi_DGP}
\left.\left.\Big/\left(36(-1+\kappa_{4}^{2}M_{p}^{2}\Omega_{D})^{2}r_{c}^{2}\right)\Bigg\}\right)\right\}\Bigg]    
\end{equation}

\begin{figure}
\centering
\includegraphics[height=2.5in,width=3.5in]{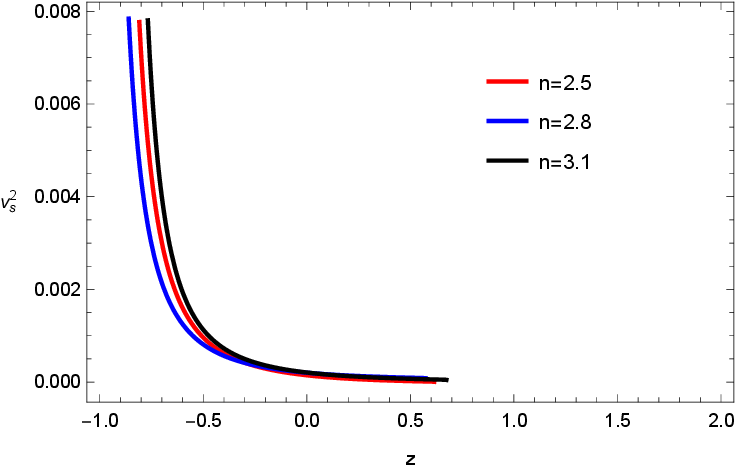}

\vspace{1 mm}

\caption{Graphical portrayal of $v_{s}^{2}$ versus $z$ for interacting NADE in DGP brane with $\Omega_{D}=0.68,~\Omega_{m_{0}}=0.28,~\epsilon=1,~\kappa_{4}=5,~H_{0}=69.25,~\delta=0.01$.}\label{G2}
\vspace{2 mm}
\end{figure}

The plots of the squared of sound speed are given in figures \ref{F9} and \ref{G2} for the interacting NADE in RS-II and DGP branes respectively. The best fit parameter values from tables \ref{tab:rsii_int} and \ref{tab:constraints} are used. Different trajectories are generated by the variation of NADE parameter $n$. We observe that here also $v_{s}^{2}>0$ for all values of $z$. These two profiles imply the classical stability of the scenario throughout the cosmological evolution.

\subsection{Statefinder Parameters}
Throughout the years, various DE models have been proposed to explain the universe's rapid expansion, leading to the difficulty of differentiating between them. To evaluate the feasibility of these models, the statefinder parameters \cite{state1, state2} are frequently utilized. The rationale for this is that the statefinder methodology may identify variations at elevated orders, even among models exhibiting closely related expansion histories. Consequently, the statefinder diagnostic methodology is most suited to differentiate across ostensibly similar models, as it incorporates higher-order adjustments in the derivatives of the scale factor. The statefinder parameters are expressed as \cite{state1, state2}
\begin{equation}
r = \frac{\dddot{a}}{aH^3},~~ \quad s = \frac{r - 1}{3(q - \tfrac{1}{2})},
\end{equation}
where $q$ represents the deceleration parameter. In terms of the Hubble and deceleration parameters, the statefinder parameters can be rewritten as
\begin{equation}
r = 2q^2 + q - \frac{\dot{q}}{H},~~~  s = \frac{r - 1}{3(q - \tfrac{1}{2})}.
\end{equation}
These parameters are dimensionless, and the associated framework is referred to as the $r-s$ plane. The paths in this plane illustrate the extent to which a specific DE model diverges from the $\Lambda$CDM benchmark. The principal regions delineated by these parameters are: $(r, s) = (1, 0)$ signifying the $\Lambda$CDM boundary, $(r, s) = (1, 1)$ representing the CDM threshold, $s > 0, r < 1$ signifies the phantom domain, whereas $s < 0, r > 1$ represents the quintessence domain.

Due to the intricate and voluminous nature of the analytical representations of the parameters $r, s$, their explicit formulations are excluded. The attributes of the cosmic $r-s$ plane are examined visually from plots for various values of $n$ in both interacting and non-interacting scenarios.

\subsubsection{Non-interacting Case}
\begin{figure}
      \centering
      \includegraphics[height=2.5in,width=3.5in]{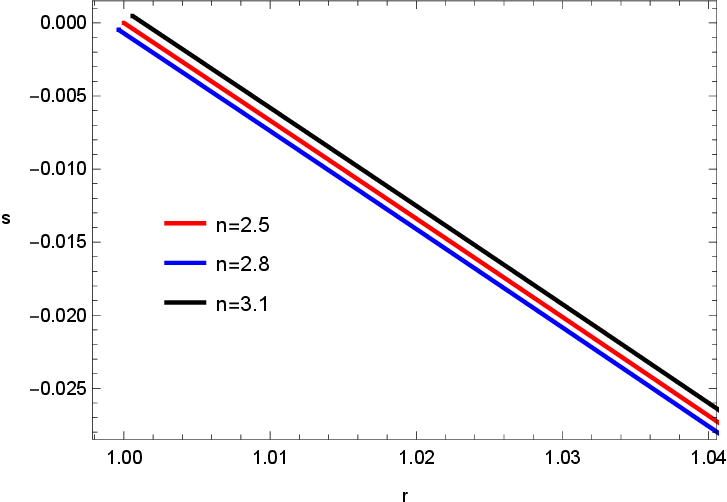}

      \vspace{1 mm}

      \caption{Graphical portrayal of $s$ versus $r$ for non-interacting RS-II with $\Omega_{D}=0.68,~\Omega_{m_{0}}=0.28,~\lambda=1.4585,~H_{0}=69.25$.}\label{F10}
      \vspace{2 mm}

\end{figure}

\begin{figure}
      \centering
      \includegraphics[height=2.5in,width=3.5in]{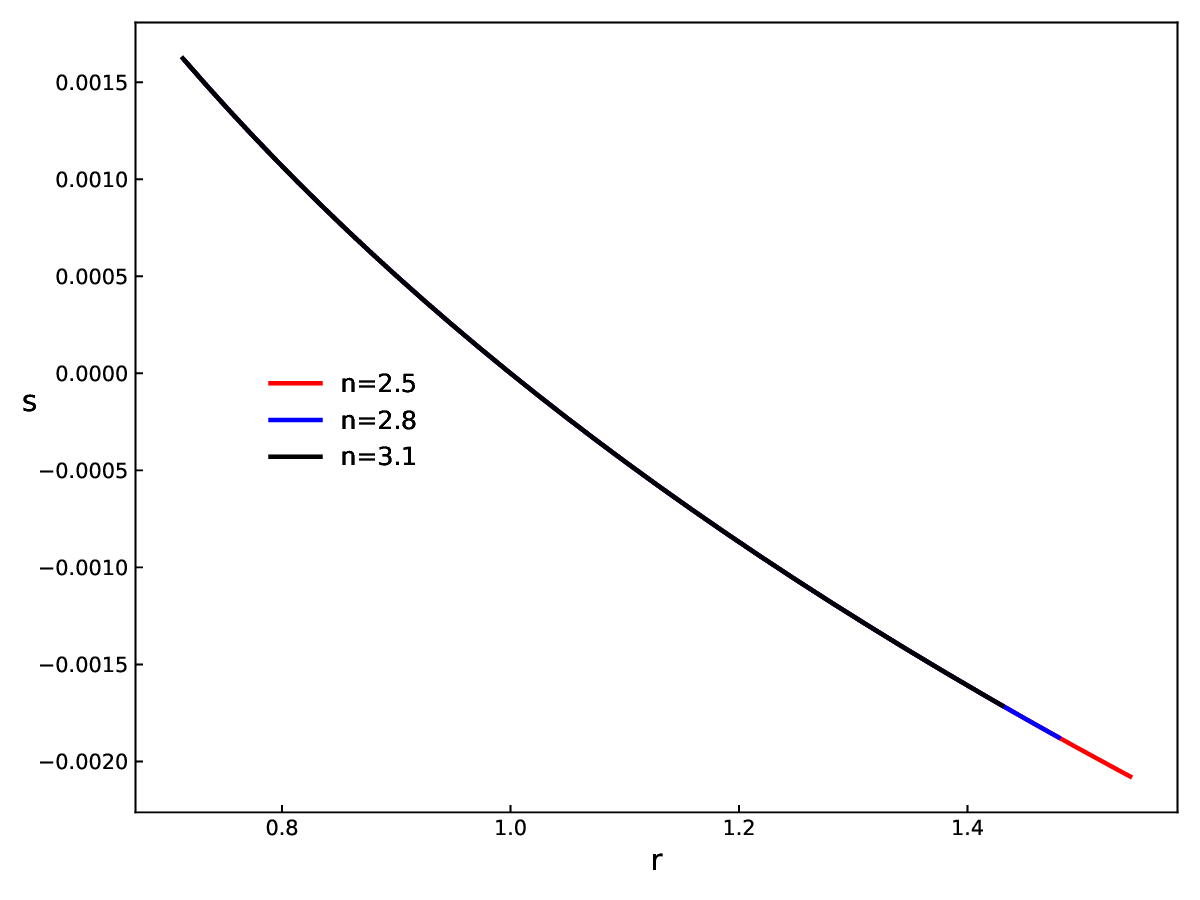}

      \vspace{1 mm}

      \caption{Graphical portrayal of $s$ versus $r$ for non-interacting DGP with $\Omega_{D}=0.68,~\Omega_{m_{0}}=0.28,~\epsilon=1,~\kappa_{4}=5,~H_{0}=69.25$.}\label{G3}
      \vspace{2 mm}

\end{figure}

The plots for the non-interacting case are given in the figures \ref{F10} and \ref{G3} for RS-II and DGP brane respectively. Here we have used the best fit parameter values from the tables \ref{tab:rsii_nonint} and \ref{tab:constraints_DGP_nonint}. We see that the trajectories of the $r-s$ plane remain in the region $s<0,~r>1$, thus showing quintessence behavior.  So it is to be noted that even the combined contributions from NADE and brane gravity, no phantom crossing is visible.

\subsubsection{Interacting Case}
\begin{figure}
      \centering
      \includegraphics[height=2.5in,width=3.5in]{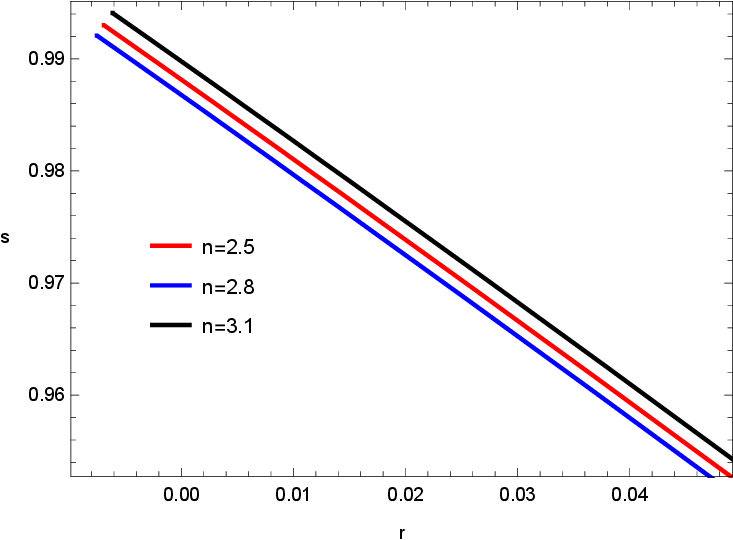}

      \vspace{1 mm}

      \caption{Graphical portrayal of $s$ versus $r$ for interacting RS-II with $\Omega_{D}=0.68,~\Omega_{m_{0}}=0.28,~\delta=0.01,~\lambda=1.4585,~H_{0}=69.25$.}\label{F12}
      \vspace{2 mm}

      \end{figure}

\begin{figure}
      \centering
      \includegraphics[height=2.5in,width=3.5in]{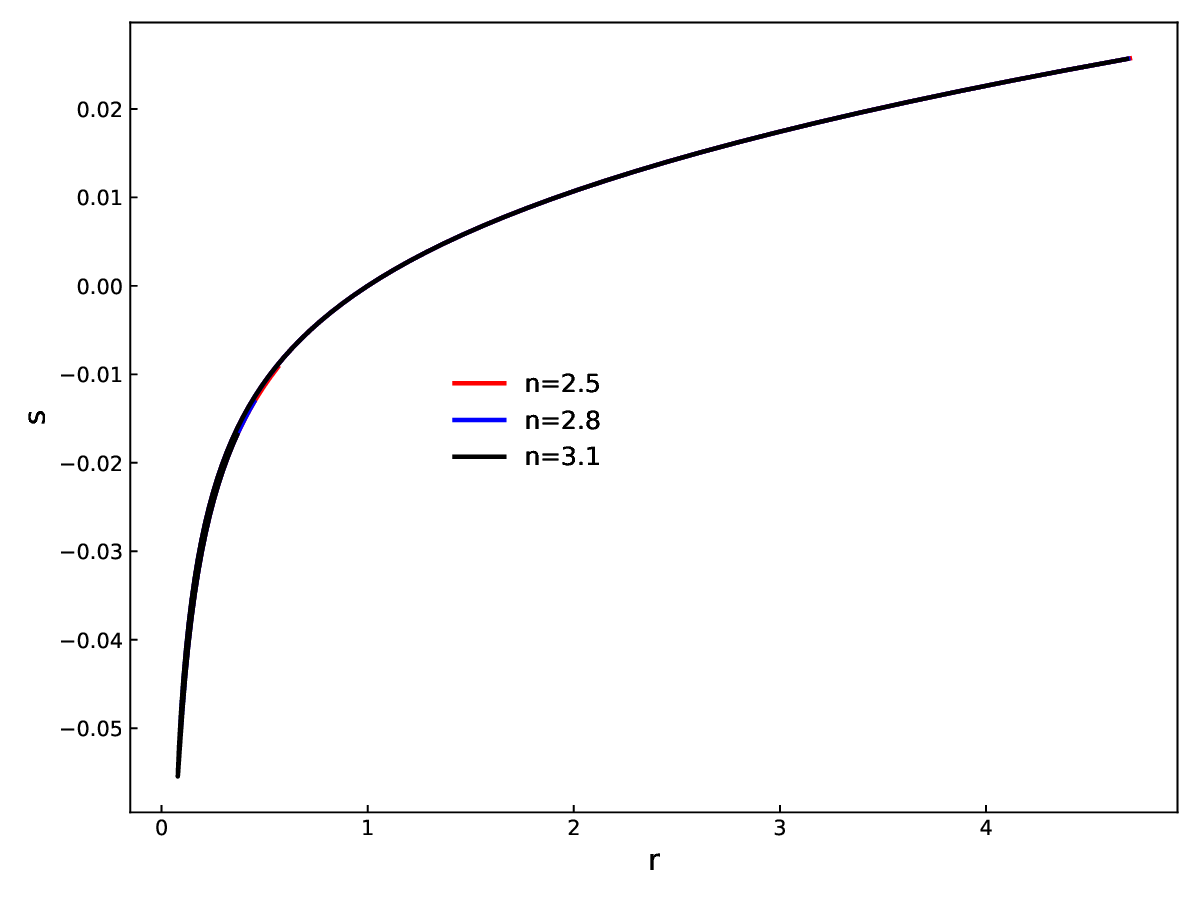}

      \vspace{1 mm}

      \caption{Graphical portrayal of $s$ versus $r$ for interacting DGP with $\Omega_{D}=0.68,~\Omega_{m_{0}}=0.28,~\epsilon=1,~\kappa_{4}=5,~H_{0}=69.25$.}\label{G4}
      \vspace{2 mm}

\end{figure}

The plots for the interacting case are given in the figures \ref{F12} and \ref{G4}, for RS-II and DGP branes respectively. Here we have used the best fit parameter values from the tables \ref{tab:rsii_int} and \ref{tab:constraints}. It is seen that the trajectories of the $r-s$ plane remain in the region $s<0,~r>1$, thus showing quintessence behavior, just like the non-interacting scenario. For both the non-interacting and the interacting scenarios, we see that the dependence of the $r-s$ trajectories on the NADE parameter is far less pronounced for the DGP brane (almost negligible) compared to the RS-II brane. We get no phantom crossing even in the presence of combined contributions from NADE and brane gravity. 

\section{Constraints on model parameters: Observational data analysis}
\label{sec:data}
Observational data analysis to constrain the parameter space of the hybrid dark energy, commonly known as Cosmic Chronometers (CC) \cite{rs14, rs15}. The covariance matrix used to estimate the likelihood is constructed following the approach described in \cite{rs16, rs17}. These data points provide direct constraints on the expansion history of the universe.
\begin{itemize}
   \item \textbf{HUBBLE Data:} We use 32 model-independent measurements of the Hubble parameter
    $H(z)$, commonly known as Cosmic Chronometers (CC) \cite{rs18, rs19}. The covariance matrix used to estimate the likelihood is constructed following the approach described in \cite{rs20, rs21}. These data points provide direct constraints on the expansion history of the universe.

    \item \textbf{BAO Data:} The BAO dataset provides constraints on the large-scale structure of the universe through measurements of the baryon acoustic feature imprinted in the matter power spectrum \cite{bao1, bao2}. We use the distance measurements from various galaxy surveys compiled in the DESI and eBOSS analyses \cite{rs22, rs23}. The observables considered include \( D_M/r_d \), \( D_H/r_d \), and \( D_V/r_d \), where \( D_M \) is the comoving angular diameter distance, \( D_H \) the Hubble distance, and \( r_d \) the comoving sound horizon at the drag epoch.
    \item \textbf{DESI Data:} The Dark Energy Spectroscopic Instrument (DESI) Release II dataset provides high-precision BAO and redshift-space distortion (RSD) measurements across a wide redshift range \cite{rs24}. These measurements significantly improve constraints on late-time cosmic acceleration and expansion rate, complementing the HUBBLE and BAO datasets.
\end{itemize}

\subsection{Statistical Analysis}
The model parameters are constrained using a Markov Chain Monte Carlo (MCMC) approach implemented via the \texttt{emcee} sampler \cite{rs25}. The posterior distributions are visualized using the \texttt{GetDist} package \cite{rs26}. For each dataset and their combinations, we compute the best-fit values and corresponding \( 1\sigma \) confidence intervals (68\% C.L.), summarized in Tables~\ref{tab:rsii_nonint}, \ref{tab:rsii_int}, \ref{tab:constraints_DGP_nonint} and \ref{tab:constraints}.

The likelihood function is defined as
\begin{equation}
    \mathcal{L} \propto \exp\left(-\frac{1}{2}\chi^2\right),
\end{equation}
where the chi-square function is given by
\begin{equation}
    \chi^2 = \Delta \mathbf{D}^T \mathbf{C}^{-1} \Delta \mathbf{D},
\end{equation}
with \(\Delta \mathbf{D} = \mathbf{D}_{\text{obs}} - \mathbf{D}_{\text{th}}\), and \(\mathbf{C}\) denoting the covariance matrix for each dataset.

\subsection{Contour Representation and Colour Scheme}

The two-dimensional confidence contours shown in the parameter constraint plots represent the joint probability distributions of the model parameters obtained from the likelihood analysis \cite{rs26}. The colour scheme of the contour plots follows the standard cosmological convention:
\begin{itemize}
    \item The \textbf{light-shaded region} corresponds to the \textbf{95\% confidence level (C.L.)}, indicating the parameter space within which the true values are expected to lie with 95\% probability.
    \item The \textbf{dark-shaded region} corresponds to the \textbf{68\% confidence level (C.L.)}, representing the $1\sigma$ range of the best-fit values.
\end{itemize}

The central black dot in each contour marks the \textbf{best-fit parameter values}, while the surrounding elliptical regions illustrate the covariance between the two fitted parameters. The darker inner contour thus represents the most probable region of parameter space consistent with the observational data, while the lighter outer contour shows the extended range allowed by the data at a higher uncertainty level. All contour plots in this analysis  ensures statistical consistency and reproducibility. The colours were chosen to maintain clarity between the 68\% and 95\% confidence regions across all polynomial bases.

\begin{figure}
    \centering
    \includegraphics[width=1\linewidth]{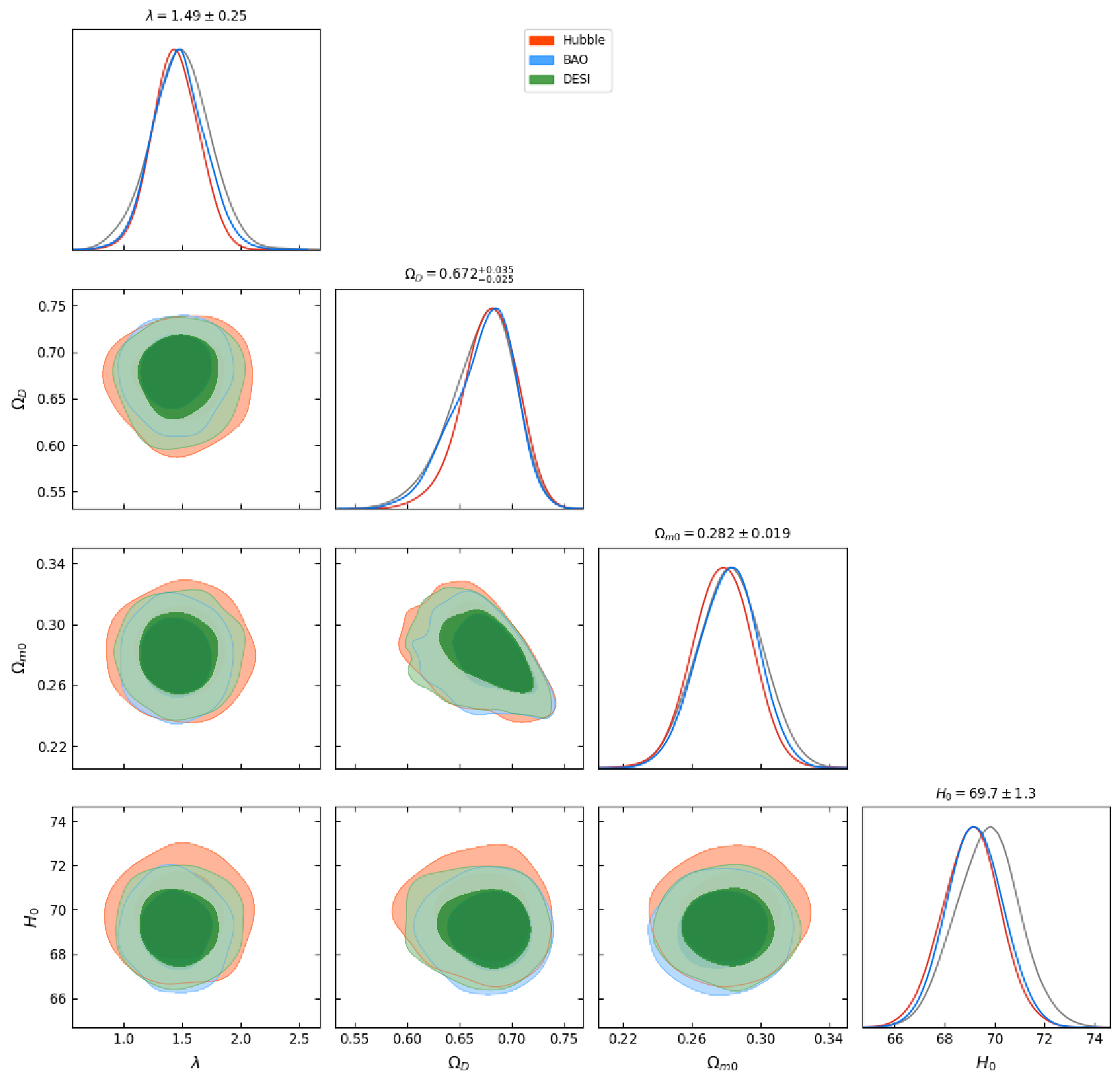}
    \caption{Joint and marginalized posterior distributions with combinations of datasets Hubble, BAO and DESI for non interacting case of RS-II brane}
    \label{fig:post_rsii_nonint_single}
\end{figure}

\begin{figure}
    \centering
    \includegraphics[width=1\linewidth]{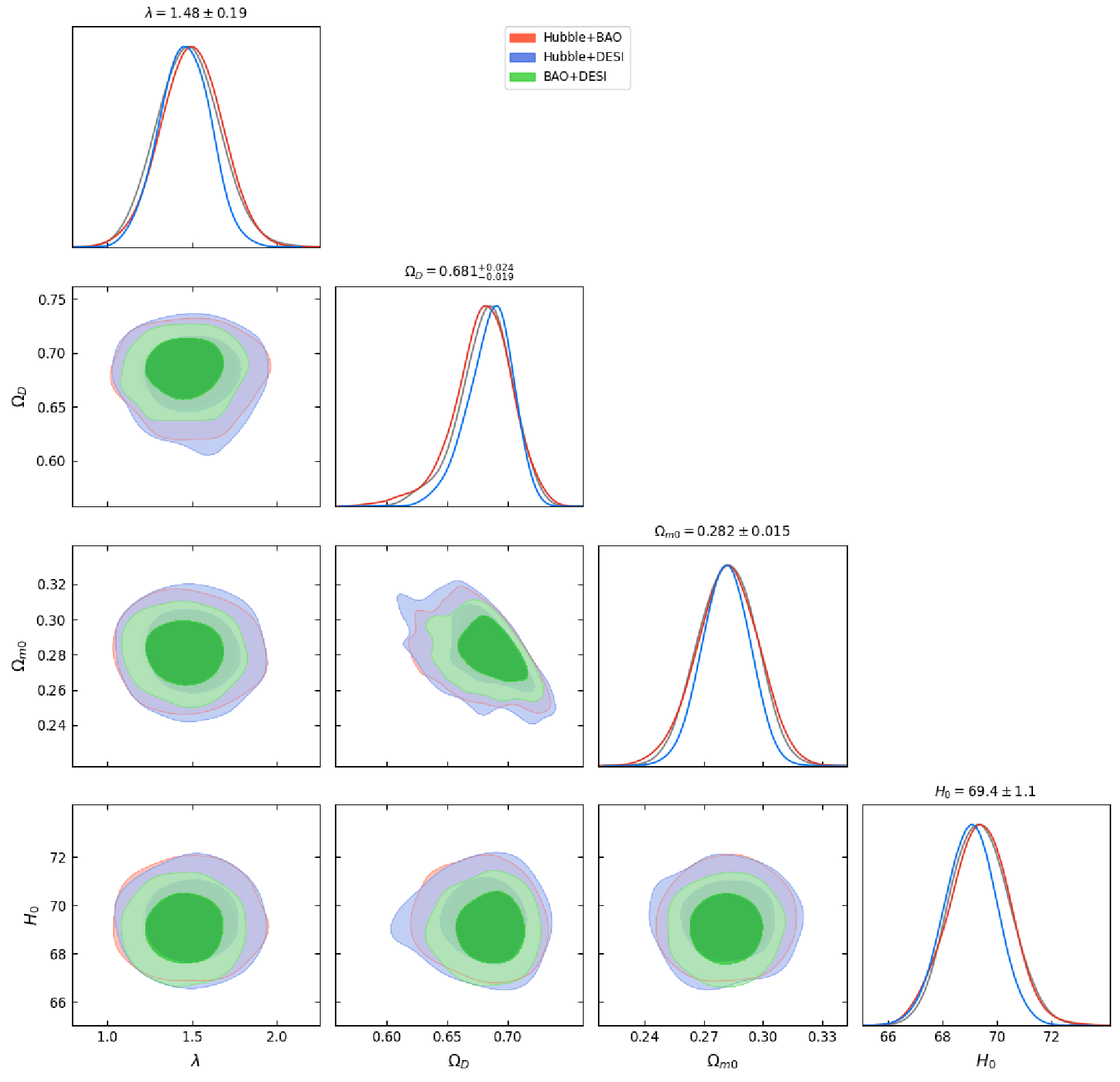}
    \caption{Joint and marginalized posterior distributions with dataset combination of Hubble+BAO, Hubble+DESI, BAO+DESI for non interacting case of RS-II brane}
    \label{fig:post_rsii_nonint_pairs}
\end{figure}

\begin{figure}
    \centering
    \includegraphics[width=1\linewidth]{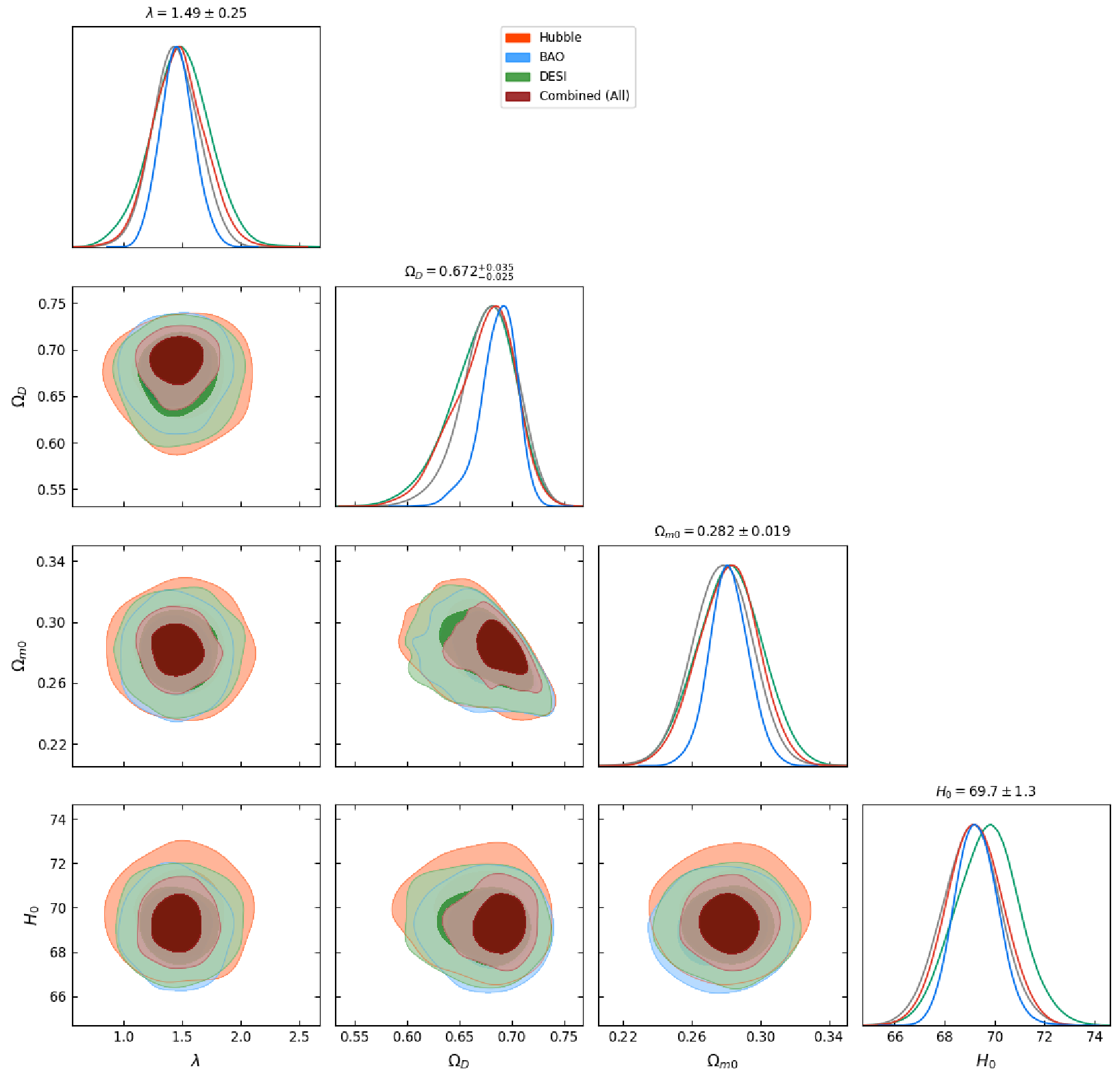}
    \caption{Joint and marginalized posterior distributions with combinations of datasets Hubble, BAO, DESI and Combined(Hubble+BAO+DESI) for non interacting case of RS-II brane}
    \label{fig:post_rsii_nonint_all}
\end{figure}

\begin{figure}
    \centering
    \includegraphics[width=1\linewidth]{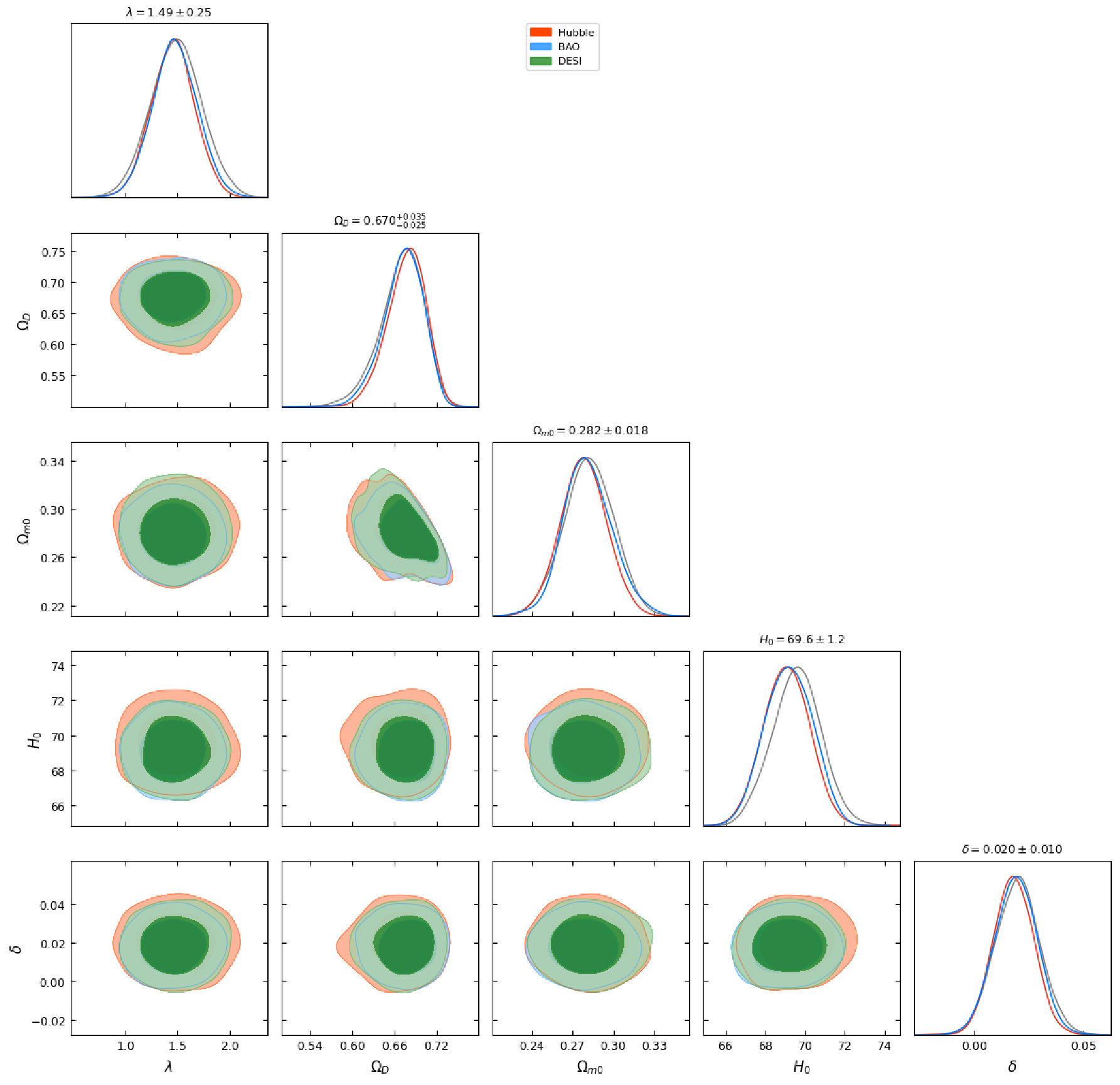}
    \caption{Joint and marginalized posterior distributions with combinations of datasets Hubble, BAO and DESI for interacting case of RS-II brane}
    \label{fig:post_rsii_int_single}
\end{figure}

\begin{figure}
    \centering
    \includegraphics[width=1\linewidth]{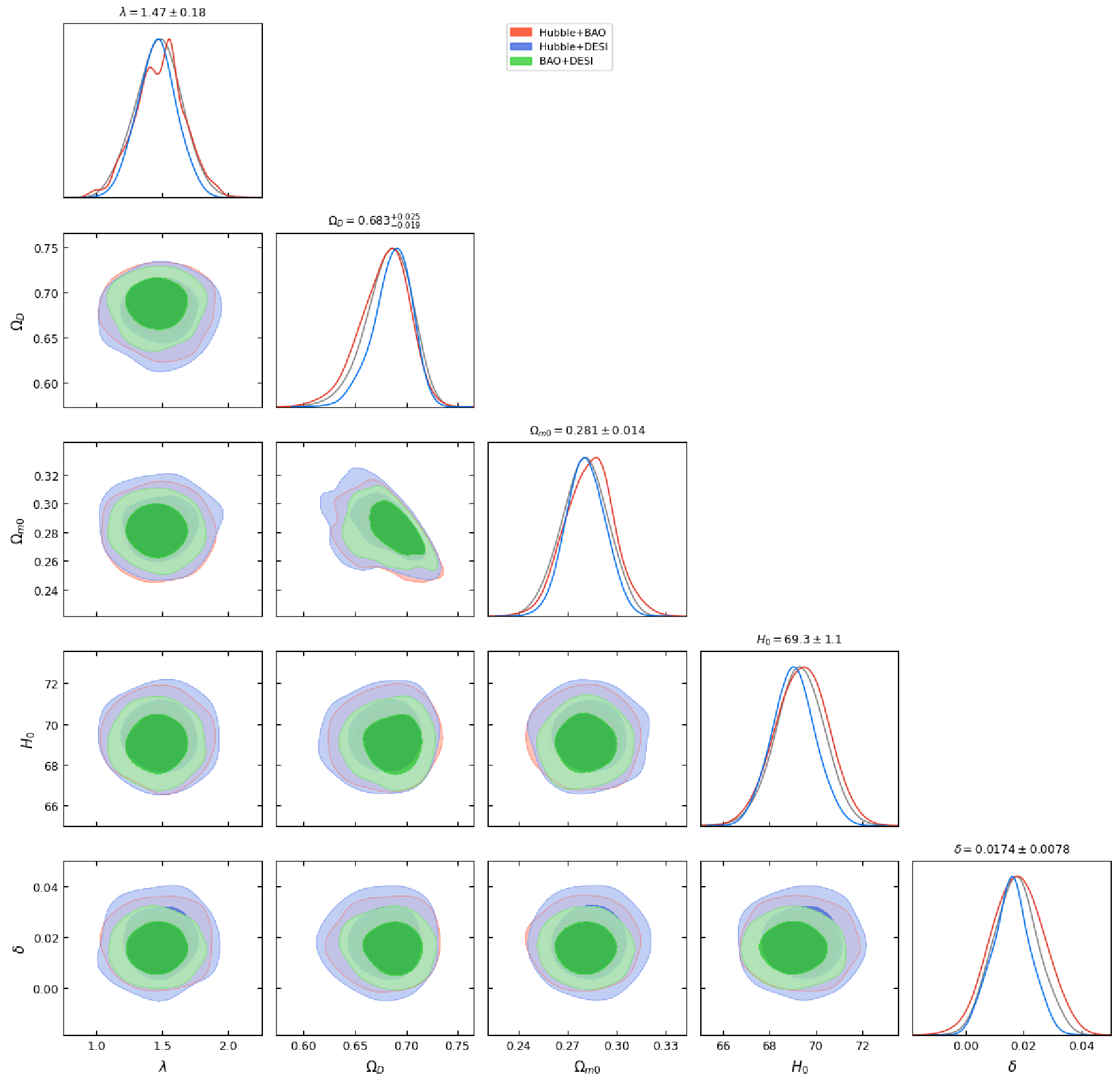}
    \caption{Joint and marginalized posterior distributions with dataset combination of Hubble+BAO, Hubble+DESI, BAO+DESI for  interacting case of RS-II brane}
    \label{fig:post_rsii_int_pairs}
\end{figure}

\begin{figure}
    \centering
    \includegraphics[width=1\linewidth]{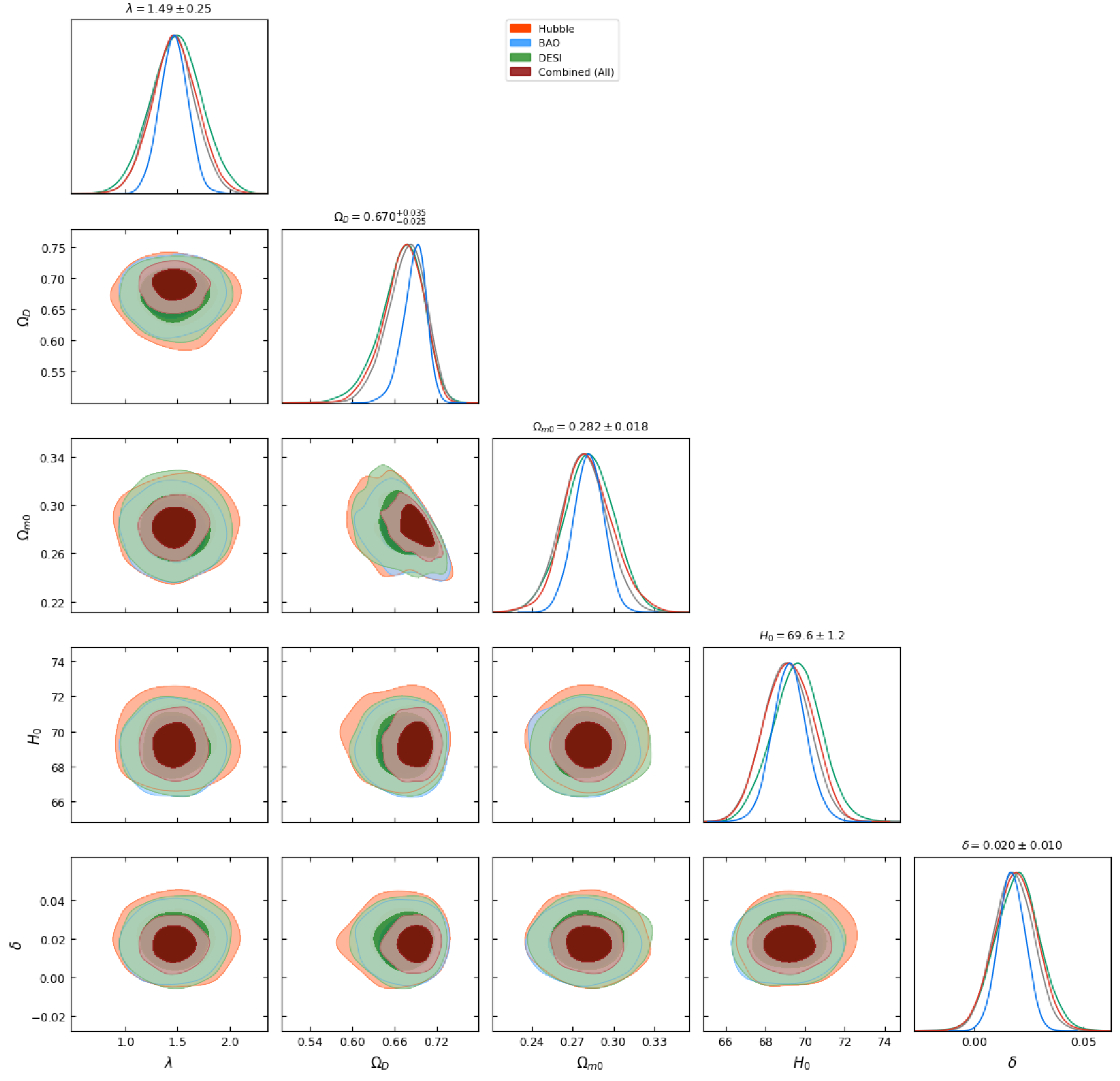}
    \caption{Joint and marginalized posterior distributions with combinations of datasets Hubble, BAO, DESI and Combined(Hubble+BAO+DESI) for  interacting case of RS-II brane}
    \label{fig:post_rsii_int_all}
\end{figure}

\begin{figure}
    \centering
    \includegraphics[width=1\linewidth]{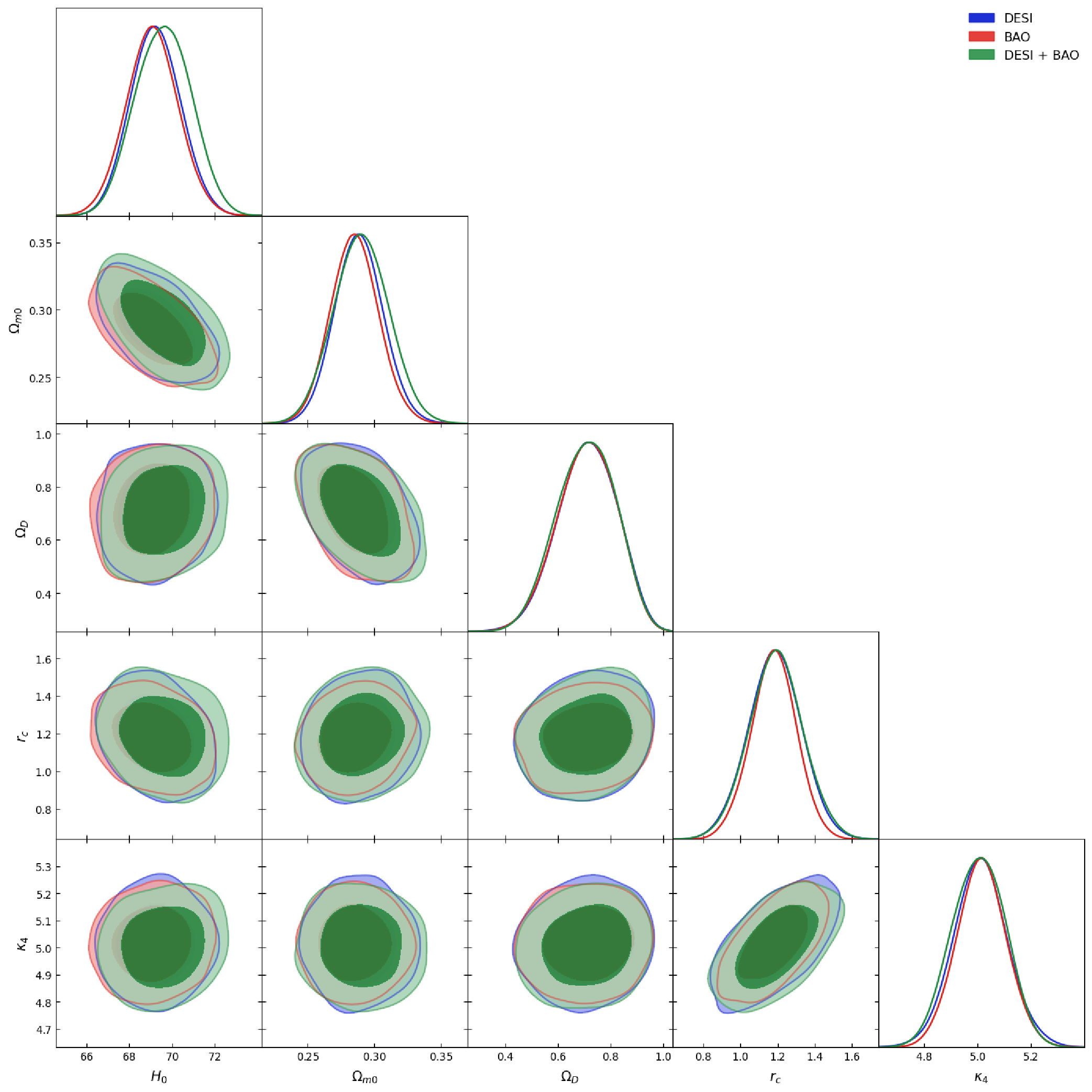}
    \caption{Joint and marginalized posterior distributions with combinations of datasets Hubble, BAO and DESI for non interacting case in DGP brane}
    \label{fig:post_dgp_nonint_single}
\end{figure}

\begin{figure}
    \centering
    \includegraphics[width=1\linewidth]{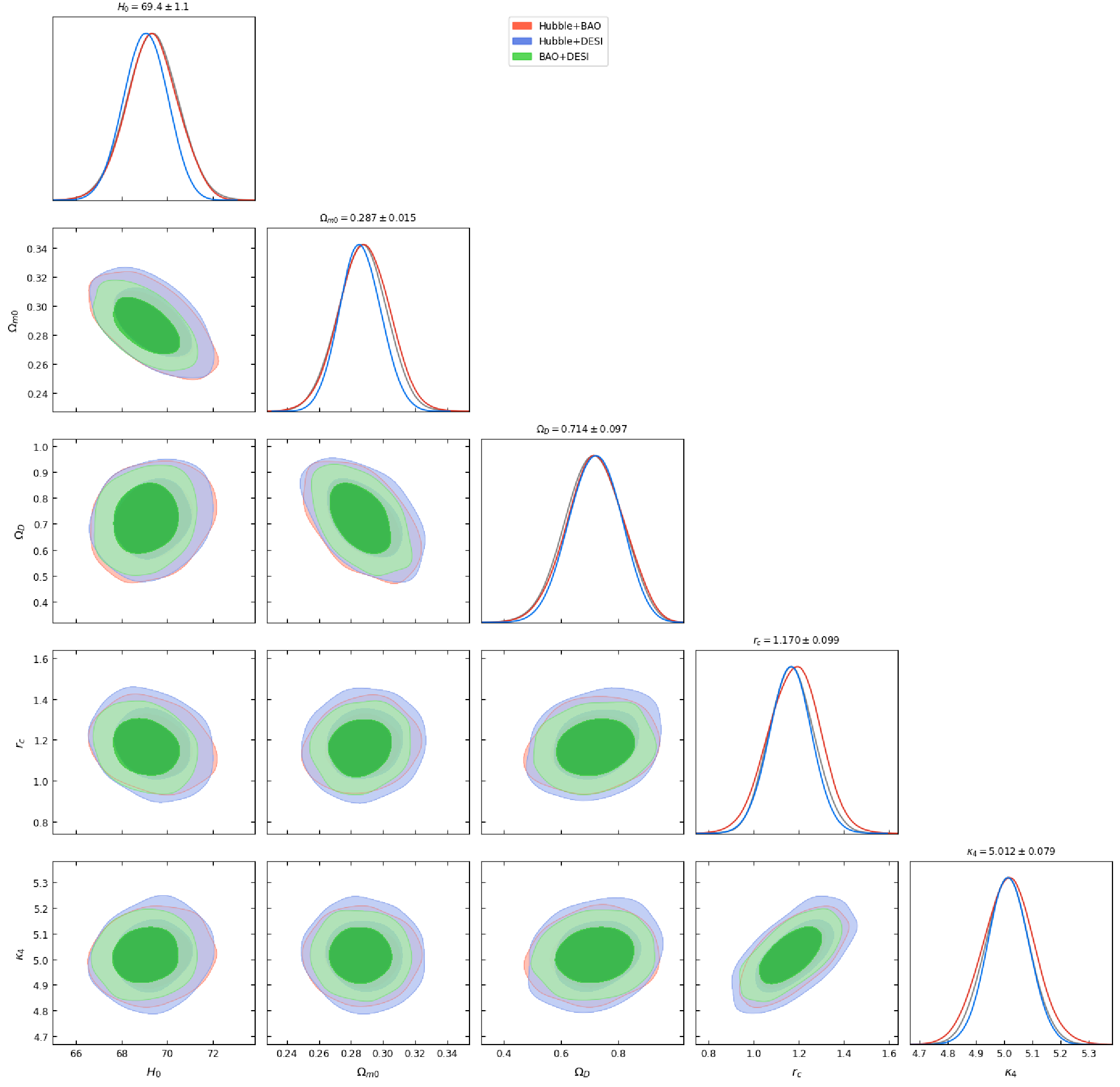}
    \caption{Joint and marginalized posterior distributions with dataset combination of Hubble+BAO, Hubble+DESI, BAO+DESI for non interacting case in DGP brane}
    \label{fig:post_dgp_nonint_pairs}
\end{figure}

\begin{figure}
    \centering
    \includegraphics[width=1\linewidth]{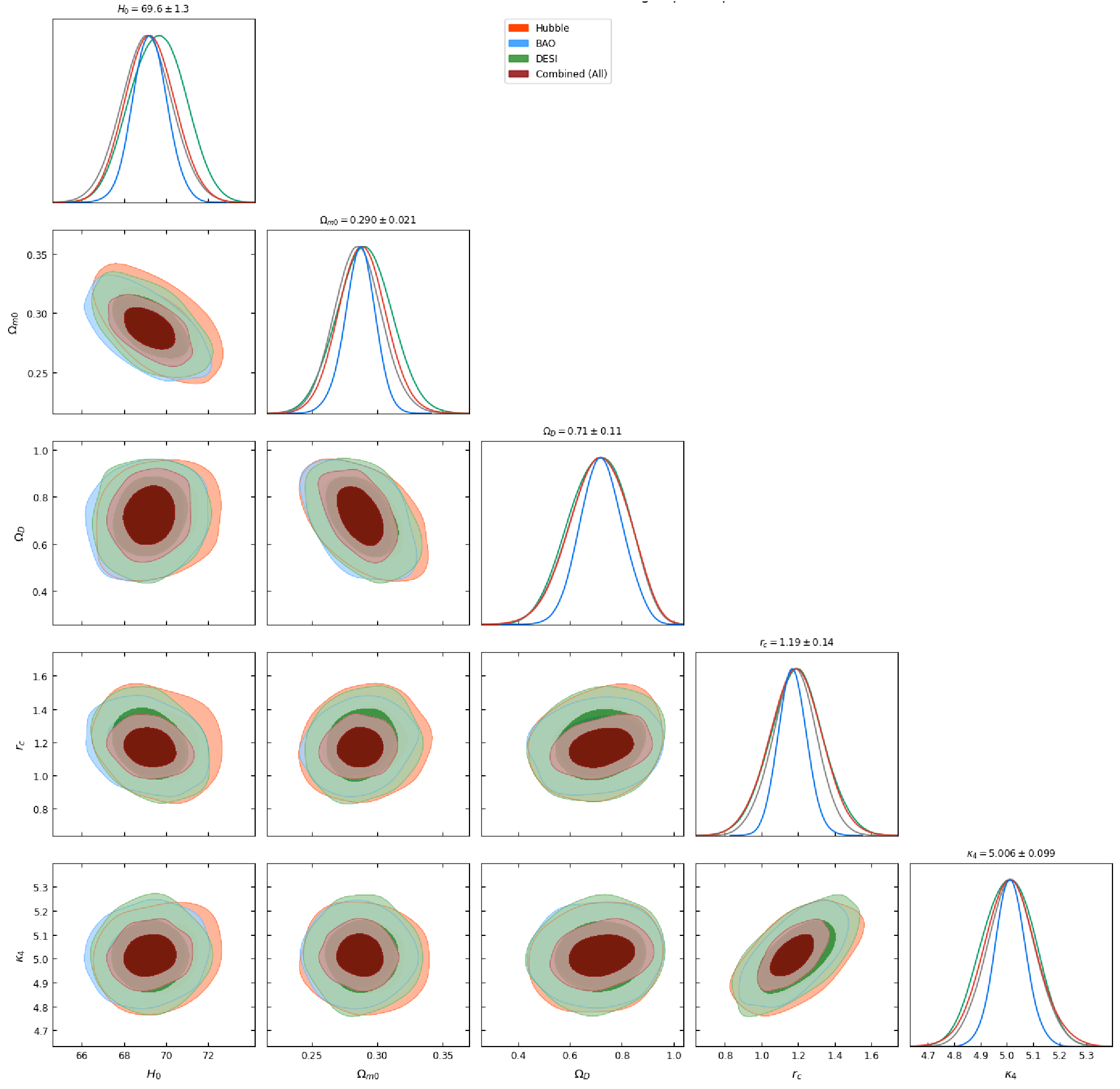}
    \caption{Joint and marginalized posterior distributions with combinations of datasets Hubble, BAO, DESI and Combined(Hubble+BAO+DESI) for non interacting case in DGP brane}
    \label{fig:post_dgp_nonint_all}
\end{figure}

\begin{figure}
    \centering
    \includegraphics[width=1\linewidth]{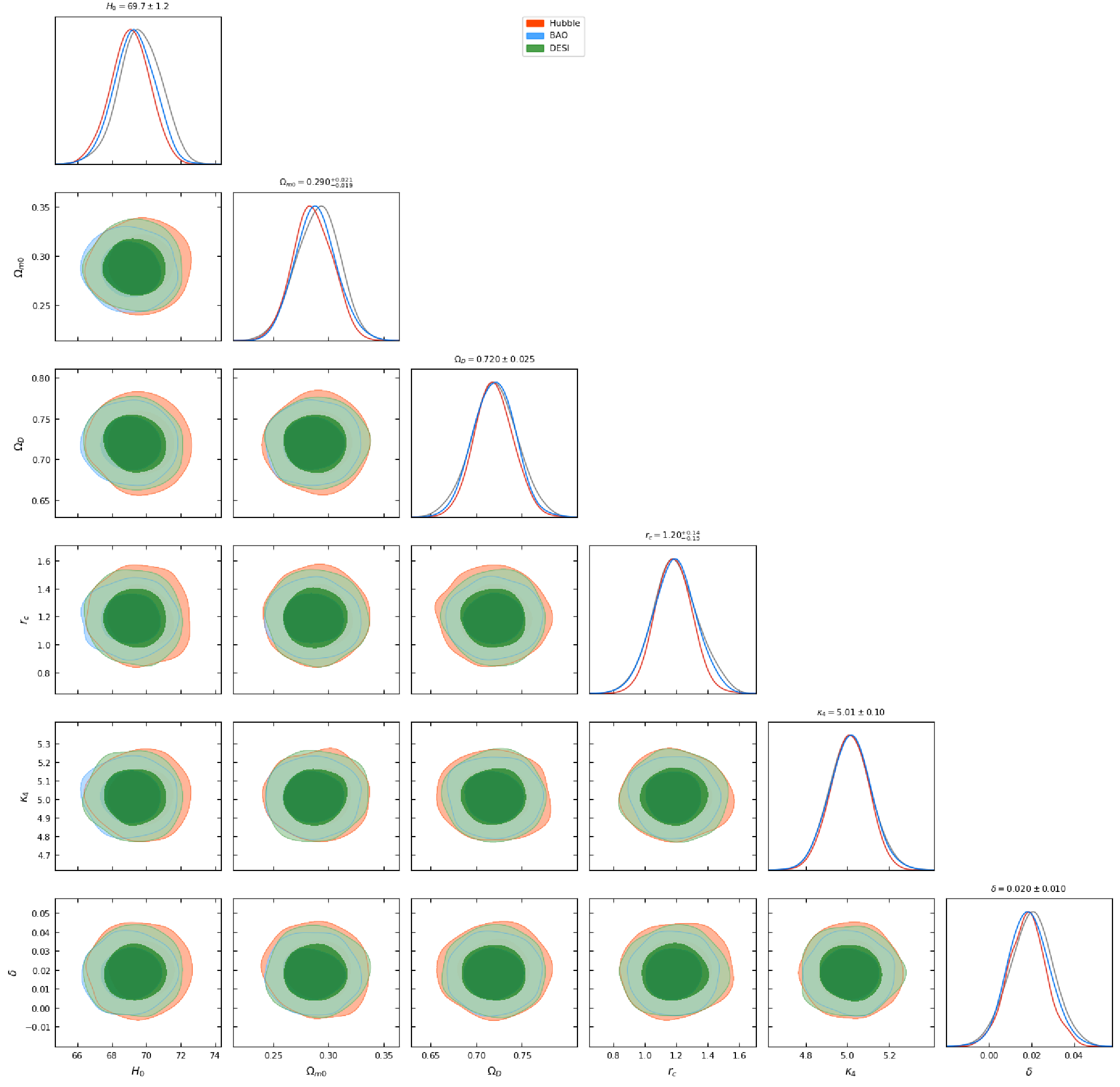}
    \caption{Joint and marginalized posterior distributions with combinations of datasets Hubble, BAO and DESI for  interacting case of DGP brane}
    \label{fig:post_dgp_int_single}
\end{figure}

\begin{figure}
    \centering
    \includegraphics[width=1\linewidth]{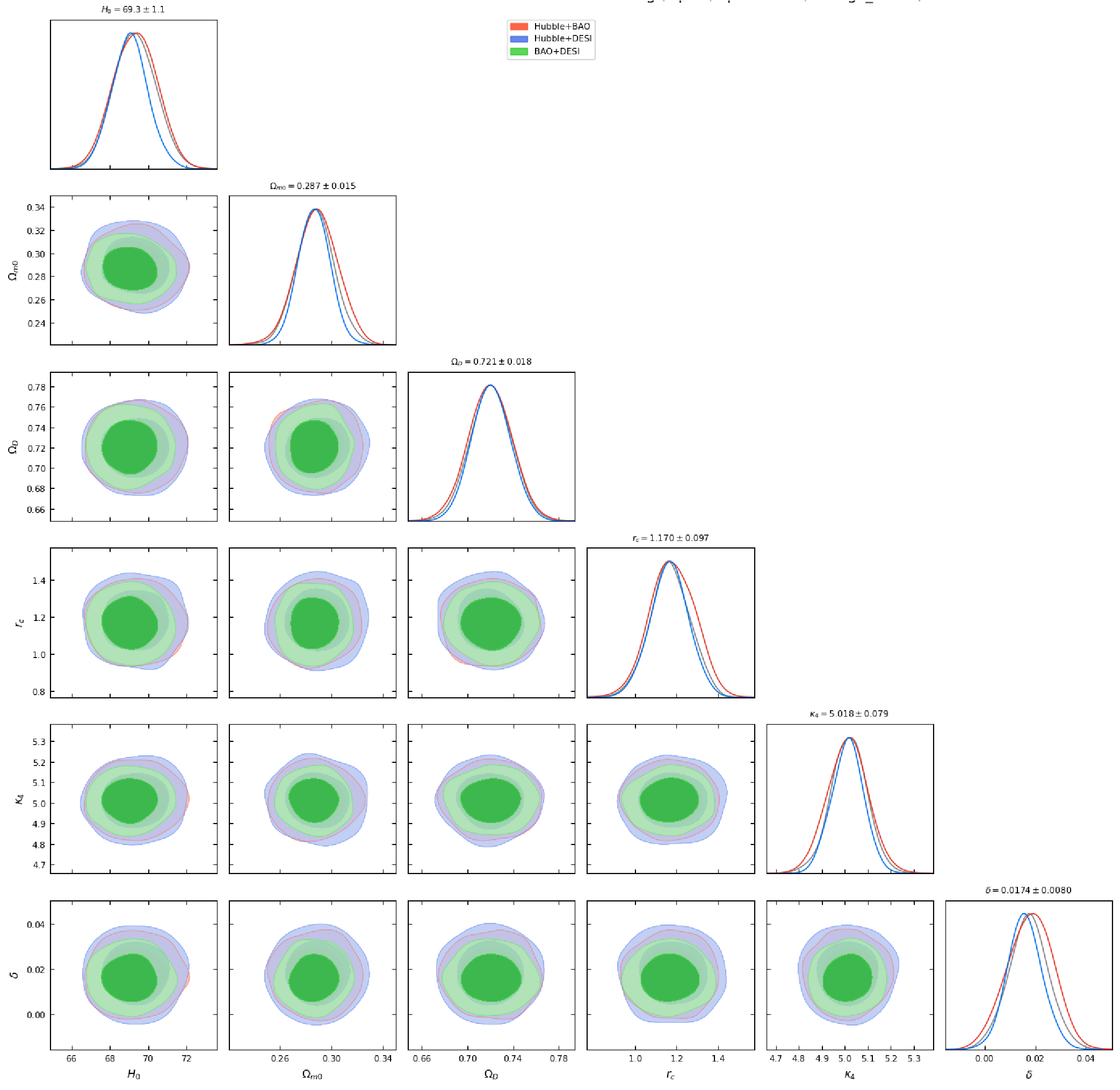}
    \caption{Joint and marginalized posterior distributions with dataset combination of Hubble+BAO, Hubble+DESI, BAO+DESI for  interacting case in DGP brane}
    \label{fig:post_dgp_int_pairs}
\end{figure}

\begin{figure}
    \centering
    \includegraphics[width=1\linewidth]{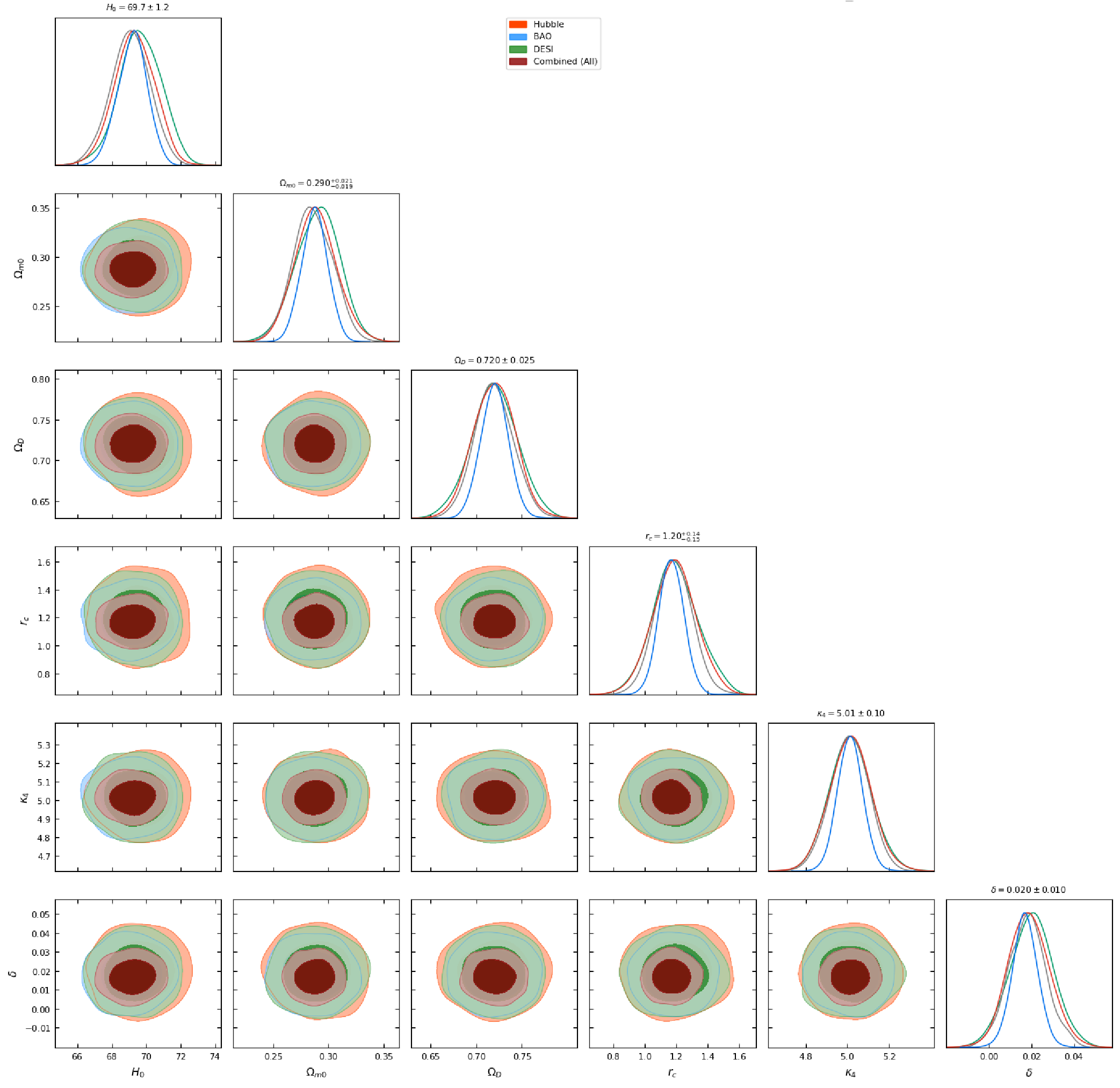}
        \caption{Joint and marginalized posterior distributions with combinations of datasets Hubble, BAO, DESI and Combined(Hubble+BAO+DESI) for  interacting case in DGP brane}
    \label{fig:post_dgp_int_all}
\end{figure}

The joint and marginalized posterior distributions for the non-interacting RS-II brane are displayed in Figs.~\ref{fig:post_rsii_nonint_single}--\ref{fig:post_rsii_nonint_all}, and the corresponding constraints are listed in Table~\ref{tab:rsii_nonint}. For the interacting RS-II brane, the posteriors are shown in Figs.~\ref{fig:post_rsii_int_single}--\ref{fig:post_rsii_int_all} with the constraints in Table~\ref{tab:rsii_int}. The results for the non-interacting DGP brane are presented in Figs.~\ref{fig:post_dgp_nonint_single}--\ref{fig:post_dgp_nonint_all} and Table~\ref{tab:constraints_DGP_nonint}, while those for the interacting DGP brane are given in Figs.~\ref{fig:post_dgp_int_single}--\ref{fig:post_dgp_int_all} and Table~\ref{tab:constraints}.



\begin{table*}[ht]
\centering
\scriptsize
\setlength{\tabcolsep}{4pt}
\renewcommand{\arraystretch}{1.15}
\caption{Parameter constraints for the non-interacting NADE model in the RS-II brane scenario obtained from different observational datasets. The quoted uncertainties correspond to the $68\%$ confidence level.}
\label{tab:rsii_nonint}

\resizebox{\textwidth}{!}{%
\begin{tabular}{lcccc}
\hline
Dataset &
$\lambda$ &
$\Omega_D$ &
$\Omega_{m0}$ &
$H_0$ (km s$^{-1}$ Mpc$^{-1}$) \\
\hline

HUBBLE &
$1.4891^{+0.2468}_{-0.2475}$ &
$0.6755^{+0.0256}_{-0.0330}$ &
$0.2824^{+0.0187}_{-0.0201}$ &
$69.7451^{+1.1879}_{-1.3850}$ \\

BAO &
$1.4468^{+0.2026}_{-0.1952}$ &
$0.6806^{+0.0249}_{-0.0265}$ &
$0.2778^{+0.0170}_{-0.0169}$ &
$69.0894^{+1.1271}_{-1.1804}$ \\

DESI &
$1.4720^{+0.2270}_{-0.2139}$ &
$0.6773^{+0.0244}_{-0.0345}$ &
$0.2820^{+0.0163}_{-0.0189}$ &
$69.2022^{+1.1536}_{-1.1121}$ \\

HUBBLE+BAO &
$1.4786^{+0.1865}_{-0.1874}$ &
$0.6832^{+0.0201}_{-0.0235}$ &
$0.2821^{+0.0149}_{-0.0158}$ &
$69.3797^{+1.0577}_{-1.1082}$ \\

HUBBLE+DESI &
$1.4998^{+0.1795}_{-0.1856}$ &
$0.6803^{+0.0211}_{-0.0239}$ &
$0.2820^{+0.0159}_{-0.0154}$ &
$69.3555^{+1.0734}_{-1.1107}$ \\

BAO+DESI &
$1.4547^{+0.1547}_{-0.1485}$ &
$0.6871^{+0.0167}_{-0.0215}$ &
$0.2814^{+0.0124}_{-0.0121}$ &
$69.0592^{+0.9297}_{-1.0129}$ \\

HUBBLE+BAO+DESI &
$1.4585^{+0.1424}_{-0.1445}$ &
$0.6895^{+0.0148}_{-0.0181}$ &
$0.2815^{+0.0119}_{-0.0107}$ &
$69.2539^{+0.8960}_{-0.8048}$ \\

\hline
\end{tabular}
}
\end{table*}


\begin{table*}[ht]
\centering
\scriptsize
\setlength{\tabcolsep}{4pt}
\renewcommand{\arraystretch}{1.15}
\caption{Parameter constraints for the interacting NADE model in the RS-II brane scenario obtained from different observational datasets. The quoted uncertainties correspond to the $68\%$ confidence level.}
\label{tab:rsii_int}

\resizebox{\textwidth}{!}{%
\begin{tabular}{lccccc}
\hline
Dataset &
$\lambda$ &
$\Omega_D$ &
$\Omega_{m0}$ &
$H_0$ (km s$^{-1}$ Mpc$^{-1}$) &
$\delta$ \\
\hline

HUBBLE &
$1.4947^{+0.2381}_{-0.2490}$ &
$0.6735^{+0.0271}_{-0.0328}$ &
$0.2818^{+0.0185}_{-0.0171}$ &
$69.6298^{+1.1380}_{-1.2652}$ &
$0.0201^{+0.0099}_{-0.0103}$ \\

BAO &
$1.4592^{+0.2060}_{-0.2233}$ &
$0.6792^{+0.0248}_{-0.0300}$ &
$0.2779^{+0.0169}_{-0.0161}$ &
$69.0731^{+1.1398}_{-1.1400}$ &
$0.0179^{+0.0095}_{-0.0092}$ \\

DESI &
$1.4704^{+0.2215}_{-0.2112}$ &
$0.6749^{+0.0275}_{-0.0290}$ &
$0.2808^{+0.0192}_{-0.0166}$ &
$69.1807^{+1.2514}_{-1.2327}$ &
$0.0192^{+0.0093}_{-0.0101}$ \\

HUBBLE+BAO &
$1.4722^{+0.1736}_{-0.1847}$ &
$0.6842^{+0.0213}_{-0.0235}$ &
$0.2808^{+0.0142}_{-0.0147}$ &
$69.3070^{+1.0308}_{-1.0067}$ &
$0.0174^{+0.0075}_{-0.0080}$ \\

HUBBLE+DESI &
$1.4911^{+0.1784}_{-0.1968}$ &
$0.6813^{+0.0207}_{-0.0275}$ &
$0.2838^{+0.0136}_{-0.0161}$ &
$69.4162^{+1.0702}_{-1.1637}$ &
$0.0180^{+0.0094}_{-0.0093}$ \\

BAO+DESI &
$1.4620^{+0.1482}_{-0.1497}$ &
$0.6880^{+0.0170}_{-0.0200}$ &
$0.2806^{+0.0128}_{-0.0119}$ &
$69.0406^{+0.9140}_{-0.9289}$ &
$0.0159^{+0.0065}_{-0.0073}$ \\

HUBBLE+BAO+DESI &
$1.4653^{+0.1401}_{-0.1372}$ &
$0.6901^{+0.0143}_{-0.0183}$ &
$0.2820^{+0.0110}_{-0.0108}$ &
$69.2336^{+0.8110}_{-0.8561}$ &
$0.0170^{+0.0065}_{-0.0058}$ \\

\hline
\end{tabular}
}
\end{table*}

\begin{table*}[ht]
\centering
\scriptsize
\setlength{\tabcolsep}{3pt}
\renewcommand{\arraystretch}{1.15}
\caption{Parameter constraints for the non-interacting NADE model in the DGP brane scenario obtained from different observational datasets. The quoted uncertainties correspond to the $68\%$ confidence level.}
\label{tab:constraints_DGP_nonint}

\resizebox{\textwidth}{!}{%
\begin{tabular}{lcccccc}
\hline
Dataset &
$H_0$ (km s$^{-1}$ Mpc$^{-1}$) &
$\Omega_{m0}$ &
$\Omega_D$ &
$r_c$ &
$\kappa_4$ \\
\hline

HUBBLE &
$69.6317^{+1.2759}_{-1.3771}$ &
$0.2903^{+0.0207}_{-0.0204}$ &
$0.7155^{+0.1100}_{-0.1184}$ &
$1.1943^{+0.1353}_{-0.1438}$ &
$5.0056^{+0.0999}_{-0.1044}$ \\

BAO &
$69.0623^{+1.2290}_{-1.2176}$ &
$0.2852^{+0.0176}_{-0.0177}$ &
$0.7146^{+0.1123}_{-0.1141}$ &
$1.1832^{+0.1142}_{-0.1248}$ &
$5.0156^{+0.0900}_{-0.0897}$ \\

DESI &
$69.2085^{+1.1988}_{-1.1264}$ &
$0.2886^{+0.0176}_{-0.0177}$ &
$0.7164^{+0.1156}_{-0.1162}$ &
$1.1904^{+0.1399}_{-0.1423}$ &
$5.0127^{+0.0952}_{-0.0983}$ \\

HUBBLE+BAO &
$69.3555^{+1.1016}_{-1.1045}$ &
$0.2873^{+0.0148}_{-0.0149}$ &
$0.7160^{+0.0986}_{-0.1003}$ &
$1.1699^{+0.1031}_{-0.1025}$ &
$5.0110^{+0.0747}_{-0.0801}$ \\

HUBBLE+DESI &
$69.3342^{+1.1304}_{-1.0777}$ &
$0.2880^{+0.0166}_{-0.0153}$ &
$0.7189^{+0.1033}_{-0.0952}$ &
$1.1848^{+0.1061}_{-0.1215}$ &
$5.0154^{+0.0897}_{-0.0938}$ \\

BAO+DESI &
$69.0677^{+0.9461}_{-0.9741}$ &
$0.2855^{+0.0133}_{-0.0116}$ &
$0.7191^{+0.0901}_{-0.0917}$ &
$1.1654^{+0.0877}_{-0.0909}$ &
$5.0150^{+0.0711}_{-0.0703}$ \\

HUBBLE+BAO+DESI &
$69.2253^{+0.7762}_{-0.7794}$ &
$0.2871^{+0.0114}_{-0.0118}$ &
$0.7207^{+0.0853}_{-0.0798}$ &
$1.1679^{+0.0786}_{-0.0753}$ &
$5.0125^{+0.0574}_{-0.0550}$ \\

\hline
\end{tabular}
}
\end{table*}
\begin{table*}[ht]
\centering
\scriptsize
\setlength{\tabcolsep}{3pt}
\renewcommand{\arraystretch}{1.15}
\caption{Parameter constraints for the interacting NADE model in the DGP brane scenario obtained from different observational datasets. The quoted uncertainties correspond to the $68\%$ confidence level.}
\label{tab:constraints}

\resizebox{\textwidth}{!}{%
\begin{tabular}{lcccccc}
\hline
Dataset &
$H_0$ (km s$^{-1}$ Mpc$^{-1}$) &
$\Omega_{m0}$ &
$\Omega_D$ &
$r_c$ &
$\kappa_4$ &
$\delta$ \\
\hline

HUBBLE &
$69.6502^{+1.2327}_{-1.1824}$ &
$0.2915^{+0.0184}_{-0.0213}$ &
$0.7199^{+0.0256}_{-0.0245}$ &
$1.1897^{+0.1535}_{-0.1346}$ &
$5.0156^{+0.0973}_{-0.1009}$ &
$0.0201^{+0.0101}_{-0.0105}$ \\

BAO &
$69.0981^{+1.0975}_{-1.1231}$ &
$0.2853^{+0.0204}_{-0.0168}$ &
$0.7191^{+0.0219}_{-0.0202}$ &
$1.1821^{+0.1197}_{-0.1141}$ &
$5.0114^{+0.0901}_{-0.0910}$ &
$0.0180^{+0.0085}_{-0.0089}$ \\

DESI &
$69.3215^{+1.1906}_{-1.1713}$ &
$0.2882^{+0.0186}_{-0.0187}$ &
$0.7201^{+0.0226}_{-0.0235}$ &
$1.1908^{+0.1362}_{-0.1439}$ &
$5.0157^{+0.0963}_{-0.1007}$ &
$0.0186^{+0.0099}_{-0.0095}$ \\

HUBBLE+BAO &
$69.3072^{+1.1130}_{-1.0678}$ &
$0.2877^{+0.0141}_{-0.0150}$ &
$0.7204^{+0.0186}_{-0.0178}$ &
$1.1672^{+0.1026}_{-0.0924}$ &
$5.0197^{+0.0740}_{-0.0765}$ &
$0.0176^{+0.0075}_{-0.0080}$ \\

HUBBLE+DESI &
$69.3731^{+1.1269}_{-1.1964}$ &
$0.2889^{+0.0167}_{-0.0161}$ &
$0.7198^{+0.0194}_{-0.0189}$ &
$1.1781^{+0.1155}_{-0.1006}$ &
$5.0159^{+0.0832}_{-0.0942}$ &
$0.0185^{+0.0086}_{-0.0098}$ \\

BAO+DESI &
$69.0632^{+0.8942}_{-0.9681}$ &
$0.2867^{+0.0117}_{-0.0123}$ &
$0.7207^{+0.0175}_{-0.0170}$ &
$1.1671^{+0.0888}_{-0.0881}$ &
$5.0145^{+0.0674}_{-0.0730}$ &
$0.0158^{+0.0068}_{-0.0067}$ \\

HUBBLE+BAO+DESI &
$69.2491^{+0.7987}_{-0.9435}$ &
$0.2877^{+0.0112}_{-0.0121}$ &
$0.7205^{+0.0149}_{-0.0156}$ &
$1.1695^{+0.0796}_{-0.0765}$ &
$5.0131^{+0.0588}_{-0.0593}$ &
$0.0168^{+0.0060}_{-0.0057}$ \\

\hline
\end{tabular}
}
\end{table*}

\section{Hubble Tension Analysis}
\label{sec:hubble_tension}

The Hubble tension stands as one of the most significant unresolved
challenges in modern cosmology \cite{rs12}. It refers to the persistent discrepancy
in the inferred value of the Hubble constant $H_0$ — the present-day
expansion rate of the universe — when derived from fundamentally
different observational probes, each interpreted within the framework of
the standard flat $\Lambda$CDM model. Two conflicting measurements lie at
the heart of this tension: early-universe determinations based on the
Cosmic Microwave Background (CMB) anisotropies measured by the Planck
satellite yield $H_0 = 67\ \mathrm{km\,s^{-1}\,Mpc^{-1}}$~\cite{rs11}, while
late-universe local distance-ladder measurements using Cepheid variables
and Type~Ia Supernovae give $H_0 = 73\
\mathrm{km\,s^{-1}\,Mpc^{-1}}$~\cite{rs13}.
The statistical significance of this disagreement has now grown to
approximately $4$--$5\sigma$, well beyond the threshold for a chance
fluctuation~\cite{rs12,rs13}. This unresolved discrepancy may point toward the presence of
unknown systematic errors in one or both measurements, or — more
intriguingly — may signal new physics beyond the standard cosmological
paradigm~\cite{rs11,rs12,rs13}.

To provide a precise and model-independent quantification of this
disagreement, we adopt the standard Gaussian tension
estimator~\cite{rs12},
\begin{equation}
  \boxed{
  T \;=\;
  \frac{\left|\,H_0^{(1)} - H_0^{(2)}\,\right|}
       {\sqrt{\,\sigma_1^2 + \sigma_2^2\,}},
  }
  \label{eq:tension_estimator}
\end{equation}
where $H_0^{(1)}$ and $H_0^{(2)}$ denote the two Hubble constant
measurements under comparison, and $\sigma_1$, $\sigma_2$ are their
respective $1\sigma$ uncertainties. The dimensionless quantity $T$
expresses the level of disagreement in units of the combined standard
deviation.

In the present work, we evaluate the Hubble tension for all four
physical scenarios considered: the RS-II brane (non-interacting and
interacting) and the DGP brane (non-interacting and interacting). In
each case, the best-fit value of $H_0$ obtained from the full combined
dataset HUBBLE+BAO+DESI — already reported in
Tables~\ref{tab:rsii_nonint}--\ref{tab:constraints} — is compared against
a $\Lambda$CDM reference value derived from the same observational
datasets. This ensures a fair, dataset-consistent comparison between the
NADE braneworld model and the standard cosmological framework.

\subsection{Hubble Tension from Best-Fit Values}
\label{subsec:bestfit_tension}

For the purpose of this analysis, the reference value of the Hubble
constant from the $\Lambda$CDM model, computed using the same
observational datasets as employed in our MCMC analysis, is
\begin{equation}
  H_0^{\,\mathrm{ref}} \;=\; 71.695\,493
  \ \mathrm{km\,s^{-1}\,Mpc^{-1}}.
  \label{eq:H0_ref}
\end{equation}
We now compute the Hubble tension $T$ for each of the four model cases
in turn.

\begin{figure}[htbp]
  \centering
  \includegraphics[width=0.72\textwidth]{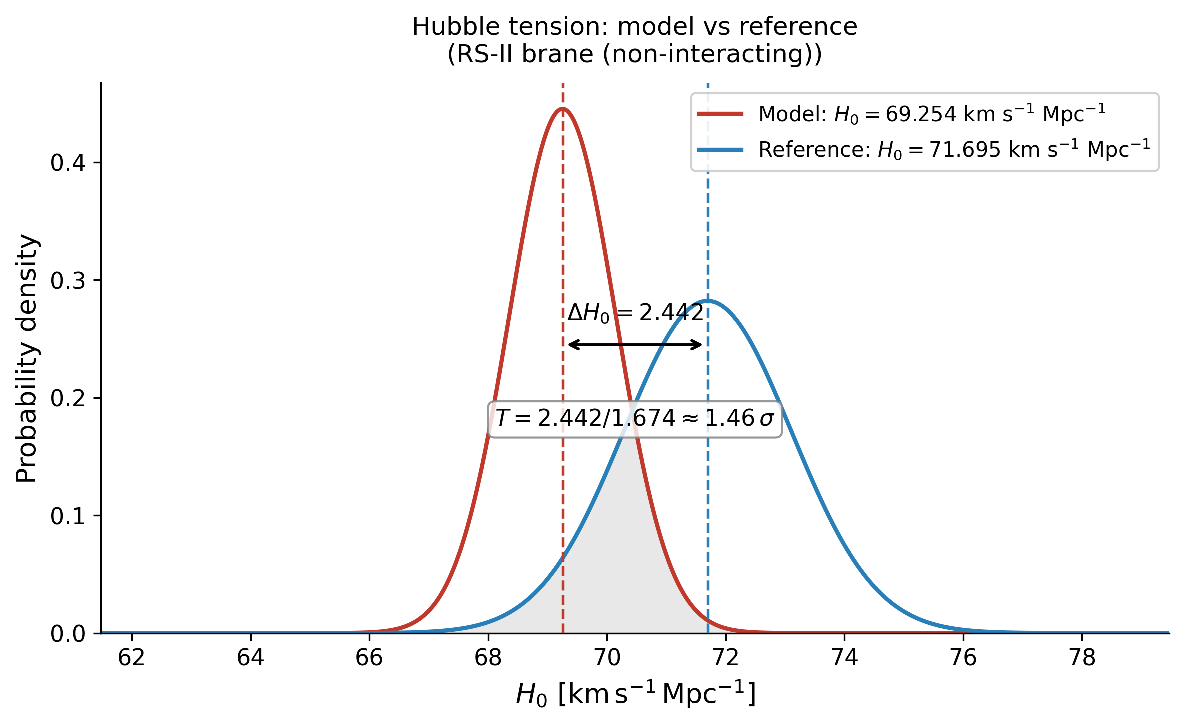}
  \caption{%
    Illustration of the Hubble tension for the \textbf{RS-II brane,
    non-interacting case}. The red (blue) Gaussian curve is centred on
    the NADE model best-fit (the $\Lambda$CDM reference) value of $H_0$.
    The double-headed arrow marks $\Delta H_0 = 2.441\
    \mathrm{km\,s^{-1}\,Mpc^{-1}}$, and the resulting tension
    $T \approx 1.45\,\sigma$ is annotated explicitly.
  }
  \label{fig:tension_RSII_nonint}
\end{figure}

\paragraph{RS-II brane — non-interacting case.}
The best-fit Hubble constant obtained from the NADE model in the
non-interacting RS-II scenario (HUBBLE+BAO+DESI combined, Table~\ref{tab:rsii_nonint}) is
\begin{equation}
  H_0^{\,\mathrm{model}} \;=\; 69.2539
  \ \mathrm{km\,s^{-1}\,Mpc^{-1}}.
  \label{eq:H0_RSII_nonint}
\end{equation}
Substituting into \cref{eq:tension_estimator}, the Hubble tension
evaluates to
\begin{equation}
  T \;\approx\;
  \frac{|\,71.695\,493 - 69.2539\,|}
       {\sqrt{\,(0.8960)^2 + (1.414)^2\,}}
  \;=\;
  \frac{2.441}{1.678}
  \;\approx\; \bm{1.45\,\sigma}.
  \label{eq:T_RSII_nonint}
\end{equation}
The corresponding Gaussian probability distributions are shown in
\cref{fig:tension_RSII_nonint}.

\begin{figure}[htbp]
  \centering
  \includegraphics[width=0.72\textwidth]{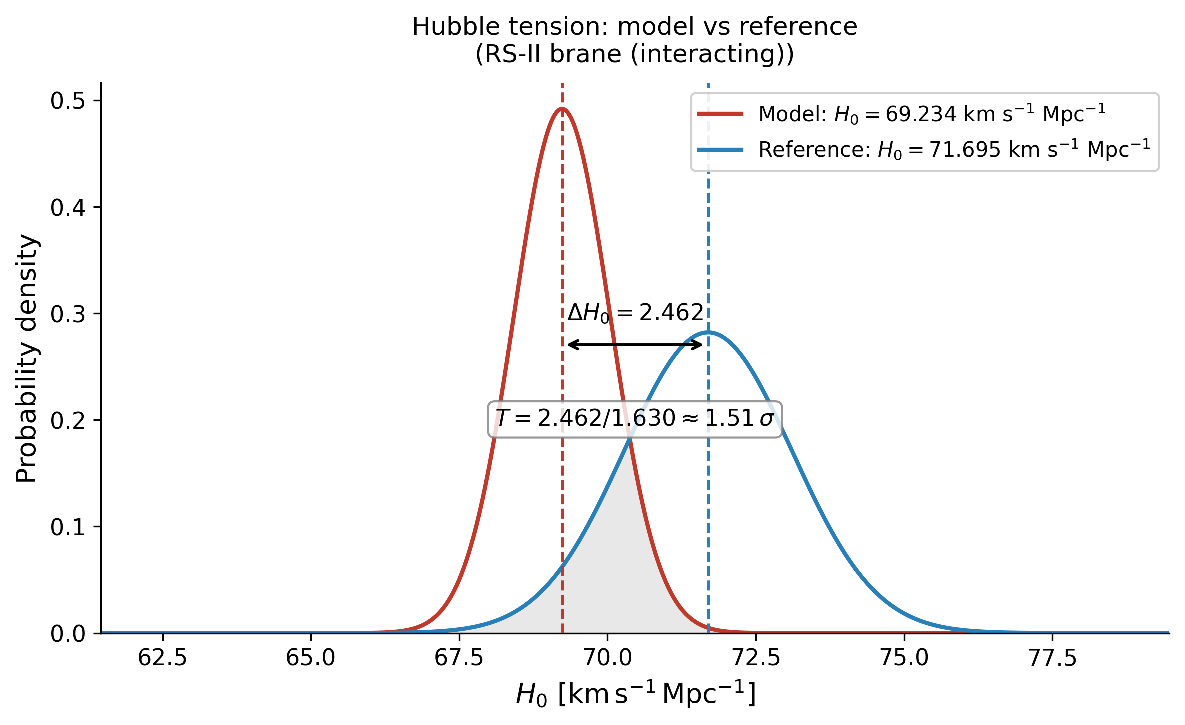}
  \caption{%
    Same as \cref{fig:tension_RSII_nonint} but for the \textbf{RS-II
    brane, interacting case}. Here $\Delta H_0 = 2.461\
    \mathrm{km\,s^{-1}\,Mpc^{-1}}$ and $T \approx 1.51\,\sigma$.
  }
  \label{fig:tension_RSII_int}
\end{figure}

\paragraph{RS-II brane — interacting case.}
For the interacting RS-II scenario (Table~\ref{tab:rsii_int}), the best-fit value is
\begin{equation}
  H_0^{\,\mathrm{model}} \;=\; 69.2336
  \ \mathrm{km\,s^{-1}\,Mpc^{-1}},
  \label{eq:H0_RSII_int}
\end{equation}
yielding a Hubble tension of
\begin{equation}
  T \;\approx\;
  \frac{|\,71.695\,493 - 69.2336\,|}
       {\sqrt{\,(0.8110)^2 + (1.414)^2\,}}
  \;=\;
  \frac{2.461}{1.635}
  \;\approx\; \bm{1.51\,\sigma}.
  \label{eq:T_RSII_int}
\end{equation}
This is illustrated in \cref{fig:tension_RSII_int}.

\begin{figure}[htbp]
  \centering
  \includegraphics[width=0.72\textwidth]{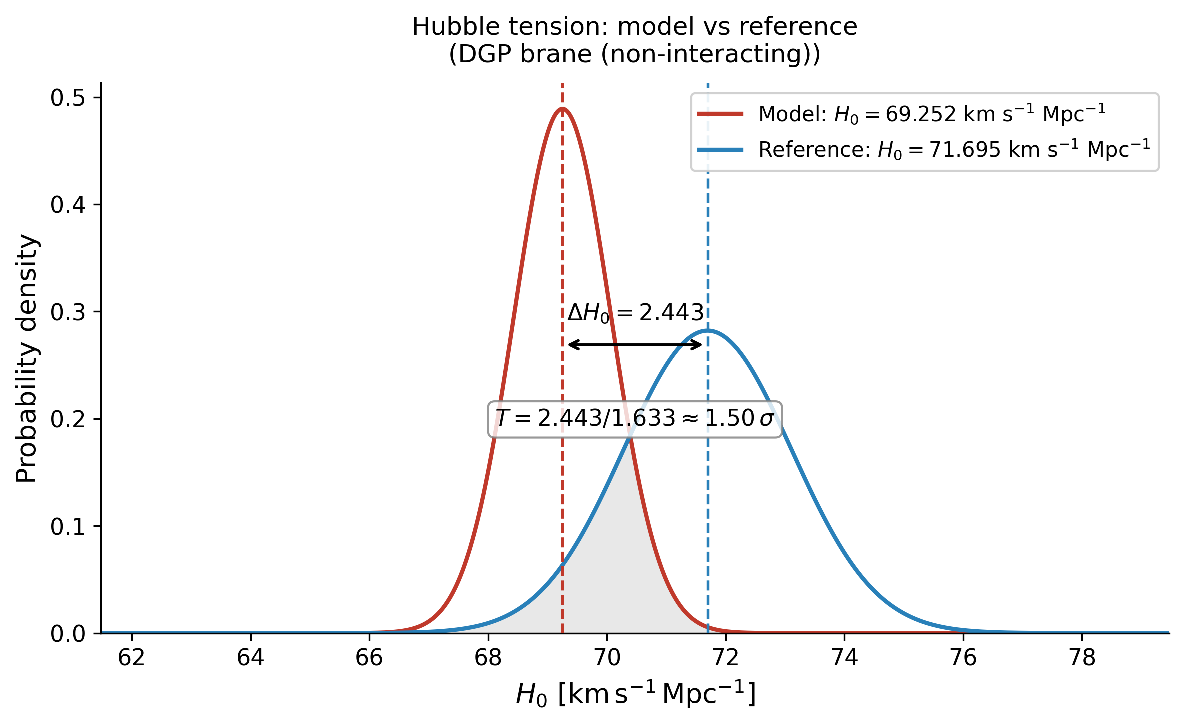}
  \caption{%
    Same as \cref{fig:tension_RSII_nonint} but for the \textbf{DGP
    brane, non-interacting case}. Here $\Delta H_0 = 2.443\
    \mathrm{km\,s^{-1}\,Mpc^{-1}}$ and $T \approx 1.49\,\sigma$.
  }
  \label{fig:tension_DGP_nonint}
\end{figure}

\paragraph{DGP brane — non-interacting case.}
In the non-interacting DGP scenario (Table~\ref{tab:constraints_DGP_nonint}), the best-fit Hubble constant is
\begin{equation}
  H_0^{\,\mathrm{model}} \;=\; 69.2523
  \ \mathrm{km\,s^{-1}\,Mpc^{-1}},
  \label{eq:H0_DGP_nonint}
\end{equation}
and the resulting tension is
\begin{equation}
  T \;\approx\;
  \frac{|\,71.695\,493 - 69.2523\,|}
       {\sqrt{\,(0.8159)^2 + (1.414)^2\,}}
  \;=\;
  \frac{2.443}{1.636}
  \;\approx\; \bm{1.49\,\sigma},
  \label{eq:T_DGP_nonint}
\end{equation}
as shown in \cref{fig:tension_DGP_nonint}.

\begin{figure}[htbp]
  \centering
  \includegraphics[width=0.72\textwidth]{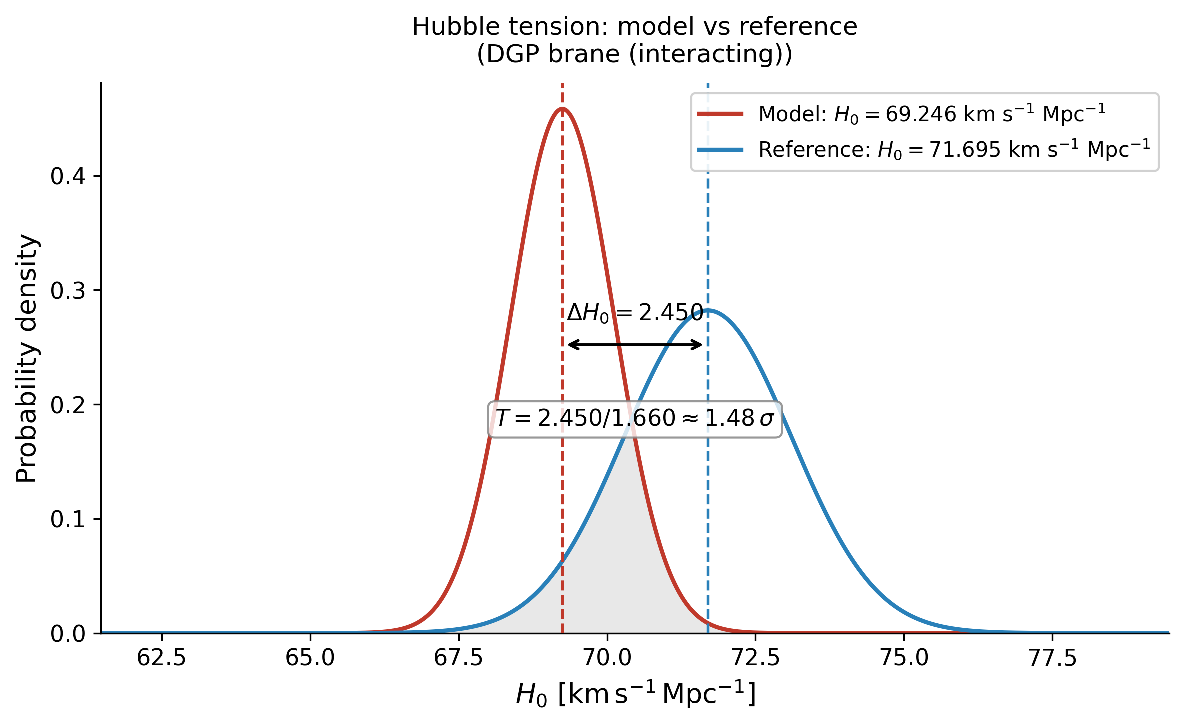}
  \caption{%
    Same as \cref{fig:tension_RSII_nonint} but for the \textbf{DGP
    brane, interacting case}. Here $\Delta H_0 = 2.449\
    \mathrm{km\,s^{-1}\,Mpc^{-1}}$ and $T \approx 1.47\,\sigma$.
  }
  \label{fig:tension_DGP_int}
\end{figure}

\paragraph{DGP brane — interacting case.}
Finally, for the interacting DGP scenario (Table~\ref{tab:constraints}), we obtain
\begin{equation}
  H_0^{\,\mathrm{model}} \;=\; 69.2459
  \ \mathrm{km\,s^{-1}\,Mpc^{-1}},
  \label{eq:H0_DGP_int}
\end{equation}
with a Hubble tension of
\begin{equation}
  T \;\approx\;
  \frac{|\,71.695\,493 - 69.2459\,|}
       {\sqrt{\,(0.8703)^2 + (1.414)^2\,}}
  \;=\;
  \frac{2.449}{1.662}
  \;\approx\; \bm{1.47\,\sigma},
  \label{eq:T_DGP_int}
\end{equation}
illustrated in \cref{fig:tension_DGP_int}.


The numerical results across all four cases are remarkably consistent,
with tensions clustering tightly in the narrow range
$1.45\sigma$--$1.51\sigma$. This consistency across two distinct
braneworld frameworks, and across both non-interacting and interacting
dark-sector scenarios, underscores the robustness of the result. While a
residual discrepancy with the $\Lambda$CDM reference value persists, it
is significantly milder than the $\sim 5\sigma$ tension reported between
early- and late-universe measurements in the standard
framework~\cite{rs11,rs12,rs13}. Crucially, the tension
obtained here is also substantially lower than the $\approx 2.3\sigma$
tension reported for the RCHDE model~\cite{Rudra2023}, demonstrating
that the higher-dimensional modifications to the Friedmann equations
encoded in the RS-II and DGP braneworld scenarios~\cite{rs1,rs2,dgp1,dgp2} provide a more
effective partial alleviation of the Hubble tension.

In all four figures below, the red curve represents the Gaussian
probability distribution centred on the NADE braneworld model best-fit
value of $H_0$, while the blue curve corresponds to the $\Lambda$CDM
reference distribution. The shaded overlap region highlights the degree
of agreement between the two distributions. The double-headed arrow
indicates the absolute difference $\Delta H_0$, and the computed tension
$T$ is annotated explicitly inside each panel.


\section{Discussion and Conclusion}
\label{sec:conclusion}
This research investigates the NADE model in the background of brane gravity, integrating the higher dimensional perspective into the evolution of dark energy. Two dfferent brane models have been used namely, the Randall–Sundrum type-II (RS-II) and Dvali–Gabadadze–Porrati (DGP) models. Utilizing conformal time as the dynamic scale the NADE model adeptly addresses the early-universe discrepancies of the initial agegraphic DE framework and exhibits complex and viable cosmological dynamics within the braneworld context. Our framework incorporates a blend of CDM and NADE, with braneworld serving as the foundational gravitational theory. The impetus arises from the pursuit of comprehending the cosmic acceleration from a higher dimensional basis. The NADE model associates DE with quantum fluctuations in spacetime and the universe's conformal age, although it does not establish a clear connection to quantum gravity. Integrating NADE inside the braneworld framework facilitates the exploration of how higher dimensional phenomena influence DE behavior, tackle cosmological singularities, enhance late-time acceleration, and provide a higher dimensional narrative of the universe's expansion.

Both interacting and non-interacting situations involving DE and CDM are probed, evaluating essential cosmological metrics including the equation of state, deceleration parameter, and statefinder parameters. The squared velocity of sound is analyzed to determine the stability of the model. For the non-interacting scenario we see that the equation of state evolves from the radiation era in the early universe to the quintessence era in the late universe for both the RS-II and DGP brane models. In fact, there is no explicit dependence of the EoS parameter on the underlying brane model. This is an interesting feature emerging from our analysis. EoS depends explicitly on the NADE parameter $n$ and the fractional DE density $\Omega_D$, while the characteristic braneworld parameters, such as the brane tension $\lambda$, the DGP crossover scale $r_c$, and the branch parameter $\epsilon$, do not appear explicitly. This result indicates that the instantaneous EoS is primarily determined by the underlying agegraphic prescription rather than by the specific form of the braneworld modification. In the interacting scenario, we see that for the RS-II brane model, we have a transition from dust to quintessence as the universe evolves, but for the DGP brane model, the trajectories cross the phantom divide and enter the phantom era in the late universe. From the study of the deceleration parameter we see that the trajectories show smooth transition from decelerated to accelerated regime around $z\approx 0.5$, which is observationally favored. The plots of the squared sound speed demonstrate classical stability of the model across the entire cosmic history. In case of the statefinder parameters we see that the trajectories are confined in the quintessence regime for both the interacting and non-interacting scenario in case of both the branes. Observational data are employed to constrain the model's free parameters. Contour plots illustrating both the combined and marginalized distributions across several datasets are produced and presented in this study. Our results suggest that the interplay of higher-dimensional brane influences on agegraphic DE provides a theoretically robust and observationally consistent description for the universe's late-time acceleration. To further investigate the phenomenological capabilities of NADE inside the braneworld cosmology framework, subsequent research could broaden this examination to include perturbative dynamics, dynamical systems, singularity analysis, and more.

\section*{Acknowledgments}
PS acknowledges Government of West Bengal Science and Technology and Biotechnology Department, for financial support to carry out Research project No.: STBT-$11012(31)/5/2024$-WBSCST SEC. PR acknowledges the Inter-University Centre for Astronomy and Astrophysics (IUCAA), Pune, India for granting a visiting associateship.

\section*{Data Availability Statement}

No data was generated or analyzed in this study.

\section*{Conflict of Interest}

There are no conflicts of interest.

\section*{Funding Statement}

There is no funding to report for this article.


\end{document}